\documentclass[preprint,showpacs,preprintnumbers,amsmath,amssymb]{revtex4-2}

\usepackage{graphicx}
\usepackage{dcolumn}
\usepackage{bm}
\usepackage{color}

\begin{document}

\newif\ifplot
\plottrue
\newcommand{\RR}[1]{[#1]}
\newcommand{\intsum}{\sum \kern -15pt \int}
\newfont{\Yfont}{cmti10 scaled 2074}
\newcommand{\Y}{\hbox{{\Yfont y}\phantom.}}
\def\O{{\cal O}}
\newcommand{\bra}[1]{\left< #1 \right| }
\newcommand{\braa}[1]{\left. \left< #1 \right| \right| }
\def\Bra#1#2{{\mbox{\vphantom{$\left< #2 \right|$}}}_{#1}
\kern -2.5pt \left< #2 \right| }
\def\Braa#1#2{{\mbox{\vphantom{$\left< #2 \right|$}}}_{#1}
\kern -2.5pt \left. \left< #2 \right| \right| }
\newcommand{\ket}[1]{\left| #1 \right> }
\newcommand{\kett}[1]{\left| \left| #1 \right> \right.}
\newcommand{\scal}[2]{\left< #1 \left| \mbox{\vphantom{$\left< #1 #2 \right|$}}
\right. #2 \right> }
\def\Scal#1#2#3{{\mbox{\vphantom{$\left<#2#3\right|$}}}_{#1}
{\left< #2 \left| \mbox{\vphantom{$\left<#2#3\right|$}}
\right. #3 \right> }}

\title{Significance of chiral three-nucleon force contact
  terms for understanding of elastic nucleon-deuteron scattering}

\author{H.~Wita{\l}a}
\email{henryk.witala@uj.edu.pl}
\affiliation{
M. Smoluchowski Institute of Physics, Jagiellonian University,
PL-30348 Krak\'ow, Poland}
\author{J.~Golak}
\affiliation{
M. Smoluchowski Institute of Physics, Jagiellonian University,
PL-30348 Krak\'ow, Poland}
\author{R.~Skibi\'nski}
\affiliation{
M. Smoluchowski Institute of Physics, Jagiellonian University,
PL-30348 Krak\'ow, Poland}

\date{Received: date / Accepted: date}

\begin{abstract}
  We investigate the importance of the three-nucleon (3N) force contact terms
  in elastic nucleon-deuteron (Nd) scattering by applying the N$^4$LO$^+$
  chiral semi-local momentum space (SMS) regularized nucleon-nucleon (NN)
  chiral potential supplemented by N$^2$LO and all subleading N$^4$LO
  three-nucleon force (3NF) contact terms. Strength parameters of
  the contact terms were obtained by least squares fitting of theoretical
  predictions to cross section and analyzing powers data at three energies
  of the impinging nucleon.  Although the N$^3$LO contributions to the
  3N force were completely neglected, the results calculated with
  the contact terms multiplied by the fitted strength parameters
  yield an improved description of the elastic Nd scattering observables
  in a wide range of incoming nucleon energies below the pion production
  threshold.
\end{abstract}


\maketitle

\section{Introduction}
\label{intro}

Since the advent of numerically exact 3N Faddeev calculations
\cite{glo96,pisa,hanover} numerous clear-cut discrepancies between
data and theoretical predictions have been found for observables in the
elastic Nd scattering and deuteron breakup reactions. Surprisingly,
the magnitudes of these discrepancies are to a large degree independent
of the dynamical ingredients used in the calculations.  They are
comparable for calculations which employ high-precision
(semi)phenomenological two-nucleon (2N) potentials supplemented
by standard models of 3N forces (3NF) and for predictions obtained
with the chiral NN and 3N N$^2$LO interactions. The low-energy analyzing
power puzzle -- a clear underestimation of the maximum of the vector
analyzing power in neutron-deuteron (nd) and proton-deuteron (pd)
elastic scattering at low incoming nucleon laboratory energies
(below $\approx 25$~MeV) -- is one of the best known cases \cite{wit2001}.
The underprediction of the elastic scattering angular distribution,
starting at $\approx 60$~MeV, in 
the region of the c.m. cross section minimum, which extends to the
backward c.m. angles and grows with the incoming nucleon
energy \cite{wit98} or a large gap at higher energies
 between the measured total
cross section for the nd interaction and theoretical
predictions \cite{abfalt98,wittot99} are further examples of
such discrepancies.  Also the breakup reaction  provides data
which remain unexplained and the most prominent case
is the cross section of the low-energy space-star geometry
in the complete nd and pd breakup \cite{wit_sst}.

In the standard approach to a 3N continuum one selects a high-precision NN
potential and augments it by some model of a 3N force, whose parameters
provide a satisfactory description of the triton binding. The 3N continuum
Faddeev equation is solved with such dynamical input and  predictions for
observables are made. For (semi)phenomenological potentials such an
attitude is from the very beginning disputable due to the inconsistency
between applied 2N and 3N interactions. The situation dramatically changed
with the availability of chiral two- and many-body forces derived
consistently in the framework of chiral perturbation
theory ($\chi$PT)
\cite{weinberg,vankolck,epel_nn_n3lo,epel6a,machl6b,epel1,epel2}.
The high precision of the description of 2N data achieved by
recent N$^4$LO$^+$ SMS NN potential of the Bochum group \cite{preinert}
together with the derivation of 3N forces up to N$^4$LO order of the
chiral expansion
\cite{epel2002,3nf_n3lo_long,3nf_n3lo_short,krebs2012,krebs2013} provided
a  solid basis  for a successful description of 3N continuum.
However, in spite of great expectations the results of  investigations
performed up  to now with the chiral 3N forces restricted to the third
order (N$^2$LO) of the chiral expansion, show that this new dynamics leads
practically  to the  same 3N data description as
 the (semi)phenomenological 2N and 3N
interactions \cite{epel2019,lenpic2021}. It means that the chiral 3NF
at N$^2$LO, which contains a 2$\pi$-exchange parameter free
component supplemented by two contact terms \cite{epel2002}, is more or
less equivalent to the commonly used Urbana IX \cite{uIX}
or TM99 \cite{TM99} 3NF's. 

The situation changed when the Pisa group published results of
calculations for elastic pd scattering below the deuteron breakup
threshold obtained with 3NF containing subleading N$^4$LO chiral contact
terms \cite{girlanda2019}.  They showed that it is possible to
correctly describe the elastic pd scattering data  together with
the  $^3$H binding energy  by augmenting  the Urbana IX 3NF with
the N$^4$LO 3NF contact terms. It indicated that very probably
the missing part of the 3N dynamics in up to now performed  3N continuum
calculations, namely omitted N$^4$LO contact terms,  could be  the
reason for difficulties in  explaining the discrepancies  mentioned above.
That offers a real prospect of even deeper understanding of 3N continuum
data up to the pion production threshold, when chiral 3N continuum
calculations with best chiral NN interaction, supplemented  by a chiral
3NF at least up to an N$^3$LO order of chiral expansion  and  including
all subleading N$^4$LO contact terms, becomes available.

The first nonvanishing contributions to the 3N force appear at N$^2$LO
\cite{vankolck,epel2002} and comprise,   in addition to
the parameter-free $2\pi$-exchange term, two contact
contributions with strength parameters $c_D$ and $c_E$
\cite{epel2002,epel_tower}.
Since the chiral 3NF acquires at N$^3$LO only parameter-free contributions,
with all N$^4$LO contact terms the 3N  Hamiltonian depends
additionally on 13 strength parameters $c_{E_i}$ of these short-range
3NF components  \cite{girlanda2011,girlanda2020a}.
Since two pairs of  N$^4$LO contact terms are identical,
the number of free parameters in the 3N Hamiltonian reduces eventually to 13.
They must be fixed by a fit to the 3N data.  This task is comparable
if not easier than in the case of the 2N system, where
the Hamiltonian required determination
of 15 free parameters \cite{preinert}.   In the 2N system fitting
parameters to 2N continuum data automatically provided the correct deuteron binding energy.  One can hope that also for the 3N system strength parameters of the 3NF contact terms obtained by fitting theoretical predictions for different  observables to 3N continuum data will lead to a 3N Hamiltonian able to reproduce the binding energies of $^3$H and $^3$He. 

Using a chiral 3NF in 3N continuum calculations  requires numerous time consuming computations  with varying strengths  of the contact terms in order to establish their values.    Fine-tuning of the 3N Hamiltonian parameters requires an extensive analysis of available 3N elastic Nd scattering and breakup data. That ambitious goal calls for a significant reduction of computer time necessary to solve the 3N Faddeev equations and to calculate the observables. Thus finding an efficient emulator for exact solutions of the 3N Faddeev equations seems to be essential and of high priority.

In Ref.~\cite{pert} we proposed such an emulator  which enabled us to reduce  significantly the required time of calculations. We  tested its efficiency  as well as ability to accurately reproduce exact solutions of 3N Faddeev equations.  In Ref.~\cite{eff_emul}   we introduced a new computational  scheme, based on the perturbative approach  of \cite{pert}, which even by far more   reduced the computer time necessary to obtain the observables in the elastic nucleon-deuteron scattering and deuteron breakup reactions at any energy, and which is well-suited for  calculations with varying strengths of the contact terms  in a chiral 3NF.

This  development of the efficient emulator for 3N continuum calculations enables us to perform an  investigation of 3N continuum with the inclusion of all 3NF contact terms up to the fifth order (N$^4$LO) of the chiral expansion. Our aim is to check whether by using the best available chiral NN potential augmented by the recently developed consistent N$^2$LO 3NF components \cite{epel2019,lenpic2021} and including  all contact terms up to N$^4$LO order of chiral expansion it is possible to fix parameters of the 3N Hamiltonian by fitting the theoretical predictions to the 3N continuum data basis.  Furthermore, we will verify if such a Hamiltonian will simultaneously reproduce the  $^3$H and $^3$He binding energies  as well as the nd doublet scattering length $^2a_{nd}$. In addition we would like to examine what impact such an Hamiltonian will have on the description of the 3N continuum, e.g. whether its use will eliminate or reduce the discrepancies mentioned earlier.

The paper is organized as follows.  In Sec.\ref{theory} for the convenience of the reader we describe the most essential points of our approach to 3N continuum calculations, especially the new emulator and very fast and efficient scheme for the computation of elastic scattering observables.  The results on importance of contributions from different N$^2$LO and N$^4$LO contact terms to numerous nd elastic scattering observables as well as on a sensitivity pattern of these contributions are shown in  Sec.\ref{importance}. In Sec.\ref{results} we determine the strengths of contact terms by fitting theoretical predictions to elastic scattering data and verify whether the established Hamiltonian leads to an improved description of Nd  elastic scattering data.  We summarize and conclude in Sec.\ref{summary}.

 \section{Theory}
 \label{theory} 

For the reader's convenience
 we briefly outline the 3N Faddeev formalism
 and the perturbative treatment of Ref.~\cite{pert}. 
 For details of the Faddeev formalism and numerical performance 
  we refer the reader to ~\cite{glo96,wit88,hub97,book}.

The 3N Hamiltonian comprises pairwise interactions 
$v_{NN} = v_{12} + v_{23} + v_{31}$
and a 3N force $V_{123}=V^{(1)}+V^{(2)}+V^{(3)}$,
where the latter is decomposed into three Faddeev components $V^{(i)}$,
symmetric in the particle labels $j, k \ne i \in \{1, 2, 3 \}$.
Since nucleons are treated as identical particles, it is possible to single out
the $(2,3)$ subsystem and use only $V^{(1)}$
in the Faddeev-type integral equation for the breakup operator $T$,
which describes Nd scattering~\cite{glo96,wit88,hub97}
\begin{eqnarray}
T\vert \phi \rangle  &=& t P \vert \phi \rangle +
(1+tG_0)V^{(1)}(1+P)\vert \phi \rangle + t P G_0 T \vert \phi \rangle \cr 
&+& 
(1+tG_0)V^{(1)}(1+P)G_0T \vert \phi \rangle \, .
\label{eq1a}
\end{eqnarray}
The initial state 
$\vert \phi \rangle = \vert \vec {q}_0 \rangle \vert \phi_d \rangle$
describes the free motion of the neutron and the deuteron 
  with the relative momentum
  $\vec {q}_0$  and contains the internal deuteron wave function
  $\vert \phi_d \rangle$.
The amplitude for elastic scattering leading to the 
 final nd state $\vert \phi ' \rangle$ is then given by~\cite{glo96,hub97}
\begin{eqnarray}
\langle \phi' \vert U \vert \phi \rangle &=& \langle \phi' 
\vert PG_0^{-1} \vert 
\phi \rangle  
 + \langle 
\phi'\vert  V^{(1)}(1+P)\vert \phi \rangle  \cr
&+& \langle \phi' \vert V^{(1)}(1+P)G_0T\vert  \phi \rangle +
\langle \phi' \vert PT \vert \phi \rangle ~,
\label{eq3}
\end{eqnarray}
while the  amplitude for the breakup reaction reads
\begin{eqnarray}
\langle  \vec p \vec q \vert U_0 \vert \phi \rangle &=&\langle 
 \vec p \vec q \vert  (1 + P)T\vert
 \phi \rangle ,
\label{eq3_br}
\end{eqnarray}
where the free  breakup channel state $\vert  \vec p \vec q \rangle $
is defined in terms of the  Jacobi (relative) momenta $\vec p$
and $\vec q$. 

We solve Eq.~(\ref{eq1a}) in the momentum-space partial-wave basis
$\vert p q \alpha \rangle$, determined by 
the magnitudes of the 
Jacobi momenta $p$ and $q$ and a set of discrete quantum numbers $\alpha$
comprising 
the 2N subsystem spin,
orbital and total angular momenta $s, l$ and $j$, as well as 
the spectator nucleon orbital
and total angular momenta with respect to the center of mass (c.m.) of the 2N
subsystem, $\lambda$ and $I$:
\begin{eqnarray}
\vert p q \alpha \rangle \equiv \vert p q (ls)j (\lambda \frac {1} {2})I (jI)J
  (t \frac {1} {2})T \rangle ~.
\label{eq4a}
\end{eqnarray}
The total 2N and spectator angular momenta $j$ and $I$ as well as isospins
$t$ and $\frac {1} {2}$, are finally 
coupled to the total angular momentum $J$ and isospin $T$ of the 3N system,
respectively.
In practice a converged solution of Eq.~(\ref{eq1a})
using partial wave decomposition
in momentum space at a given energy $E$ requires taking all 3N partial wave
states up to the 2N angular momentum $j_{max}=5$ 
and the 3N angular momentum $J_{max}=\frac{25}{2}$, with
the 3N force acting up to the 3N total
angular momentum $J=7/2$. The number of resulting
partial waves for given $J$ (equal to the number of coupled integral
equations in two continuous
variables $p$ and $q$)
amounts to $142$. The required computer time to get one solution on a 
personal computer is about
 few hours. In the case when such calculations have to be performed for a
big number of varying 3NF parameters, time restrictions
become prohibitive.

The perturbative approach proposed
in Ref.~\cite{pert} and \cite{eff_emul}  
leads to a significant reduction of the required computational time.
 It relies on the fact that it is possible to apply a perturbative
 approach in order to include the contact terms in 3N continuum calculations.
 Let us consider a chiral 3NF $V^{(1)}$ at a given order of
 chiral expansion with
variable strengths of its contact  terms.
 The contact terms are restricted to 
 small 3N total angular momenta and
 to only few partial wave states for a given total 3N angular momentum 
 $J$ and parity $\pi$ \cite{epel2002,epel_tower}.  
 We split the $V^{(1)}$  into a parameter-free term $V(\theta_0)$
 and a sum of $N$ contact terms $c_i \Delta V_i$ 
   with strengths $c_i$: 
\begin{eqnarray}
  V^{(1)}  &=&V(\theta_0)   +   \Delta V(\theta)
  = V(\theta_0)   +  \sum_{i=1}^N c_i \Delta V_i  ~,
\label{eq4}
\end{eqnarray}
with $\theta_0=(c_i=0,i=1,\dots,N)$ and $\theta=(c_i,i=1,\dots,N)$ being 
the sets
of contact terms strength values, 
 for which we would like to find solution of
 Eq.~(\ref{eq1a}).

We divide the 3N partial wave states into two sets: $\beta$ and
the remaining one, $\alpha$. The $\beta$ set is defined by nonvanishing
matrix elements
of $\Delta V(\theta)$.
 Introducing $T(\theta_0)$ and $\Delta T(\theta)$ such that 
$T \equiv T(\theta)=T(\theta_0) + \Delta T(\theta)$,
 and using the fact that $\Delta V(\theta)$ has nonvanishing elements only
 for channels $\vert \beta \rangle$, one gets from Eq.~(\ref{eq1a})
  two separate sets of equations for
 $\langle \alpha \vert T (\theta_0) \vert \phi \rangle$ and
  $\langle \alpha \vert \Delta T (\theta) \vert \phi \rangle$
  (Eqs.(9) and (10) in \cite{pert} or Eqs.(6) and (7) in \cite{eff_emul}). 
  The first equations in sets (6) and (7) of Ref. \cite{eff_emul} 
  are the
 Faddeev equations (\ref{eq1a}) for $T(\theta_0)$. 
 The second equation in the set (7) for
 $\langle \beta \vert \Delta T (\theta) \vert \phi \rangle$ can be solved
 within the set of channels $\vert \beta \rangle$ only.
Since $\Delta V(\theta)$ is  small,
it is possible to neglect the term 
 $ \langle \beta \vert (1+tG_0) \Delta V(\theta)
 (1+P)G_0 \Delta T(\theta)   \vert \phi \rangle $ in the kernel
 and arrive at the following integral equation for
  $\langle \beta \vert \Delta T (\theta) \vert \phi \rangle$:
\begin{eqnarray}
  \langle \beta \vert \Delta T (\theta) \vert \phi \rangle &=&
 \langle \beta \vert (1+tG_0) \Delta V(\theta) (1+P)\vert \phi  \rangle \cr
 &+& \langle \beta \vert (1+tG_0) \Delta V(\theta) (1+P)G_0 T(\theta_0)
 \vert \phi  \rangle \cr
 &+& \langle \beta \vert (1+tG_0) V (\theta_0) 
 (1+P)G_0 \Delta T(\theta)   \vert \phi \rangle  \cr
 &+& \langle \beta \vert t P G_0 \Delta T(\theta) \vert \phi \rangle ~. 
\label{eq11}
\end{eqnarray}
That equation permits one to transfer the linear
dependence  on the strengths $c_i$ from 
 the $\Delta V(\theta)$ on  
 the  $\Delta T(\theta)$. Namely, let
 $ \langle \beta \vert \Delta T_i  \vert \phi \rangle $ be a solution
 of Eq.(\ref{eq11}) for a
 set $\theta_i=(c_i=1, c_{k \ne i}=0)$:
 \begin{eqnarray}
  \langle \beta \vert \Delta T_i \vert \phi \rangle &\equiv&
 \langle \beta \vert (1+tG_0) \Delta V_i (1+P)\vert \phi  \rangle \cr
 &+& \langle \beta \vert (1+tG_0) \Delta V_i (1+P)G_0 T(\theta_0)
 \vert \phi  \rangle \cr
 &+& \langle \beta \vert (1+tG_0) V (\theta_0) 
 (1+P)G_0 \Delta T_i   \vert \phi \rangle  \cr
 &+& \langle \beta \vert t P G_0 \Delta T_i \vert \phi \rangle ~.
\label{eq12}
\end{eqnarray}
then the solution of Eq.~(\ref{eq11}) is given by:
\begin{eqnarray}
  \langle \beta \vert  \Delta T(\theta) \vert \phi \rangle &=&
  \sum_{i=1}^N c_i  \langle \beta \vert  \Delta T_i  \vert \phi \rangle ~.
\label{eq14}
\end{eqnarray}

In this way at a given  energy the computation of  observables in the elastic
Nd scattering and deuteron breakup reaction for  any combination of strengths
$c_i$ of contact terms is reduced to solving  once  $N+1$  
  Faddeev equations: one equation for $T(\theta_0)$
 and N equations for $\Delta T_i$. In the first step,
 solution for
 $ \langle \alpha  (\beta) \vert T (\theta_0) \vert \phi \rangle $  is found. 
 Then  Eq.~(\ref{eq12}) is solved for
$ \langle \beta \vert \Delta T_i \vert \phi \rangle $, from which  the 
$ \langle \alpha \vert \Delta T_i  \vert \phi \rangle $ is calculated by:
\begin{eqnarray}
\langle \alpha \vert \Delta T_i \vert \phi \rangle &=&
\langle \alpha \vert t P G_0  \sum_{\beta} \int_{p'q'}
\vert p' q' \beta \rangle \langle p' q' \beta \vert
\Delta T_i \vert \phi \rangle \cr 
&+& \langle \alpha \vert (1+tG_0) V (\theta_0) (1+P)G_0 \cr
&&\sum_{\beta} \int_{p'q'}  \vert p'q' \beta \rangle
\langle p' q' \beta \vert   \Delta T_i  \vert \phi \rangle ~.
\label{eq15}
\end{eqnarray}

The computations described above need to be done only once and then for any combination
of the strengths $c_i$
$\langle \alpha ( \beta) \vert T(\theta=(c_i,i=1,\dots,N) ) \vert \phi \rangle $
is obtained by trivial summation:
\begin{eqnarray}
  \langle \alpha \vert T (\theta) \vert \phi \rangle &=&
  \langle \alpha \vert T (\theta_0) \vert \phi \rangle
  + \sum_i c_i \langle \alpha \vert \Delta T_i \vert \phi \rangle  \cr
  \langle \beta \vert T (\theta) \vert \phi \rangle &=&
  \langle \beta \vert T (\theta_0) \vert \phi \rangle + \sum_i c_i
  \langle \beta \vert \Delta T_i \vert \phi \rangle ~.
\label{eq16}
\end{eqnarray}

For a calculation of elastic scattering observables 
the required sum of the second and the  third  term in Eq.~(\ref{eq3})
is obtained by:
\begin{eqnarray}
 && \langle  \alpha \vert  V^{(1)}(\theta)(1+P)\vert \phi \rangle
  + \langle  \alpha \vert V^{(1)}(\theta)(1+P)G_0T(\theta)
  \vert  \phi \rangle = \cr
 && \langle  \alpha \vert  V(\theta_0)(1+P)\vert \phi \rangle
  +  \langle  \alpha \vert  V(\theta_0)(1+P)G_0 T(\theta_0)
  \vert \phi \rangle \cr
&&+ \sum_i c_i [ \langle  \alpha \vert \Delta V_i(1+P) \vert \phi \rangle
    +  \langle  \alpha \vert \Delta V_i(1+P)G_0 T(\theta_0)
    \vert \phi \rangle \cr
&&    +  \langle  \alpha \vert V(\theta_0) (1+P)G_0 \Delta T_i 
  \vert \phi \rangle ] \cr  
&& + \sum_{i,k} c_i c_k  \langle  \alpha \vert \Delta V_i (1+P)G_0 \Delta T_k
  \vert \phi \rangle ~.
\label{eq17}
\end{eqnarray}

This computational scheme constitutes the improved emulator of Ref.~\cite{eff_emul}.
The efficiency of the fitting procedure based on this emulator can be further increased
because 
the 
elastic scattering $\langle \phi' \vert U \vert \phi \rangle $ and
breakup $\langle  \vec p \vec q \vert U_0 \vert \phi \rangle$ transition
amplitudes are linear in the matrix elements $\langle p q \alpha \vert T \vert \phi \rangle$.
Therefore also the final transition matrix elements 
are linked to the strengths $c_i$ in the same way as shown in Eqs.(\ref{eq16}) and (\ref{eq17}).
It allows us at a given energy to perform only once all required
 interpolations, integrations over Jacobi momenta, as well as 
  summations over the partial waves, total angular momenta and parities, to gain
  the contributions to the transition amplitudes, which are independent
  from the actual values of strengths $c_i$. 
  Finally, the  transition amplitudes for any particular
  set of strengths $c_i$ are obtained 
  from the same simple relations as in
  Eqs.(\ref{eq16}) and (\ref{eq17}). It reduces radically the required
  time to compute observables and permits to get observables for hundreds of
  strengths combinations in the blink of an eye. 
  An additional beneficial feature of that emulator, which makes it
  especially well-suited for optimization purposes, is the simple dependence
  of the transition amplitudes on the strengths $c_i$, enabling fast and easy
  access to the gradient with respect to strength parameters for any observable.

\section{Importance of contact terms in elastic nd scattering}
\label{importance}

Equipped with the new emulator we investigate the significance of
the 3NF contact terms for understanding the Nd elastic scattering. To this end we
take the state-of-the-art chiral SMS N$^4$LO$^+$
NN potential of the Bochum group \cite{preinert} with the regularization
parameter $\Lambda = 450$~MeV, combined with the N$^2$LO chiral 3NF \cite{epel2002}
and supplemented by all subleading N$^4$LO 3NF contact
terms \cite{girlanda2011,girlanda2020a}.
Such a Hamiltonian comprises altogether 15 short range contributions to 3NF, two from
N$^2$LO \cite{vankolck,epel2002} with the strengths $D$ and $E$, and
thirteen from N$^4$LO  \cite{girlanda2020a} with the
strengths $E_i, i=1,\dots,13$. To all these contact terms we applied the same
nonlocal Gaussian regulator defined in Eq.~(13) of Ref.\cite{epel_tower} with the
cutoff parameter $\Lambda =450$~MeV. The strengths $D$, $E$, and  $E_i$ can be
expressed by dimensionless coefficients $c_D$, $c_E$, and $c_{E_i}$
according to \cite{epel_tower}:
\begin{eqnarray}
D = \frac {c_{D}}  {F^2_{\pi} \Lambda_{\chi}},~~   
E = \frac {c_{E}}  {F^4_{\pi} \Lambda_{\chi}},~~
E_i = \frac {c_{E_i}}  {F^4_{\pi} \Lambda^3_{\chi}} ~, 
\label{eq_18}
\end{eqnarray}
where $F_{\pi}= 92.4$~MeV is the pion decay constant and $\Lambda_{\chi}=700$~MeV. 

In order to explore the role of contact terms in 3N continuum,
partial wave decomposition (PWD)
of these 3NF components must be performed in the momentum space. It has been 
 done in a standard way \cite{book,glo-coon}. The corresponding
expressions for $D$ and $E$ terms can be found in Ref.\cite{epel2002} and for
$E_1$ and $E_7$ terms in Ref.\cite{epel_tower}.
 For the remaining N$^4$LO contact terms the choice 
 of the Faddeev component and its PWD
 are given in the  Appendix. It turned out that
 PWD for the $E_{9}$ and  $E_{11}$ terms yields the same result.
 The same was found for the
 $E_{10}$ and $E_{12}$ terms. Thus only 11 out of 13 N$^4$LO contact
 terms are independent.

In view of the large number of contributing short-range terms 
one can ask the question, whether 
it is at all possible to find the unambiguous magnitudes of all these strengths 
using available Nd elastic scattering data.
Only if the answer is affirmative, valuable predictions based on the 
resulting 3N Hamiltonian can be obtained.

Before we answer this pivotal question, let us
consider the patterns, according to which short range 3NF terms
contribute to different Nd elastic scattering observables.
In particular it should be examined if some terms are more
important than others for a specific class of observables, and how
a pattern of sensitivity to different 3NF contact terms is changing
 with energy.

To that end we performed the 3N continuum Faddeev calculations 
at five laboratory energies of the
incoming neutron $E=10$, $70$, $135$, $190$, and $250$~MeV
using the dynamical input defined above and our emulator.

The selected energies
cover the range of interesting discrepancies between theory and data
 mentioned in the introduction. To find out the pattern of sensitivity
to a particular short range component we calculated at these energies
all elastic nd scattering observables 
adding consecutively to the parameter free 3NF part $V(\theta_0)$
only one component with a strength $c_{E_i}$ varied between
$c_{E_i}=-2.0$ up to  $c_{E_i}=+2.0$.  The set of elastic scattering observables
(55 in total) comprised the differential cross section,
nucleon vector and deuteron vector and tensor analyzing powers, spin
correlation coefficients, nucleon to nucleon, nucleon to deuteron,
deuteron to nucleon and deuteron to deuteron spin transfer coefficients.
For each observable we studied angular variations of predictions themselves
 as well as 
 a quantity $\Delta$ which shows the sign and magnitude of the percent deviation
  from the prediction with the parameter free 3NF term ($V(\theta_0)$)
  induced by that specific 3NF component,
 averaged over all c.m. angles $\theta_k$. Specifically, for a particular
 observable Obs, and for 
 only one short range term with the strength $c_j$ active,
 ($c_j$ is one of the strength $c_D$, $c_E$, or $c_{E_i}, i=1,\dots,10,13$), 
 the quantity $\Delta$ is defined by:
\begin{eqnarray}
  \Delta \equiv \Delta(c_j) = \frac {1} {N_{\theta}} \sum_{\theta_k} \frac {Obs(c_j,\theta_k)
    -Obs(\theta_0,\theta_k)}     {Obs(\theta_0,\theta_k)} \times 100 \% ~,
\label{eq_19}
\end{eqnarray}
with $N_{\theta}=73$ and step of $\theta_k$ equal $2.5^\circ$. 

\begin{figure}    
\includegraphics[scale=0.55,clip=true]{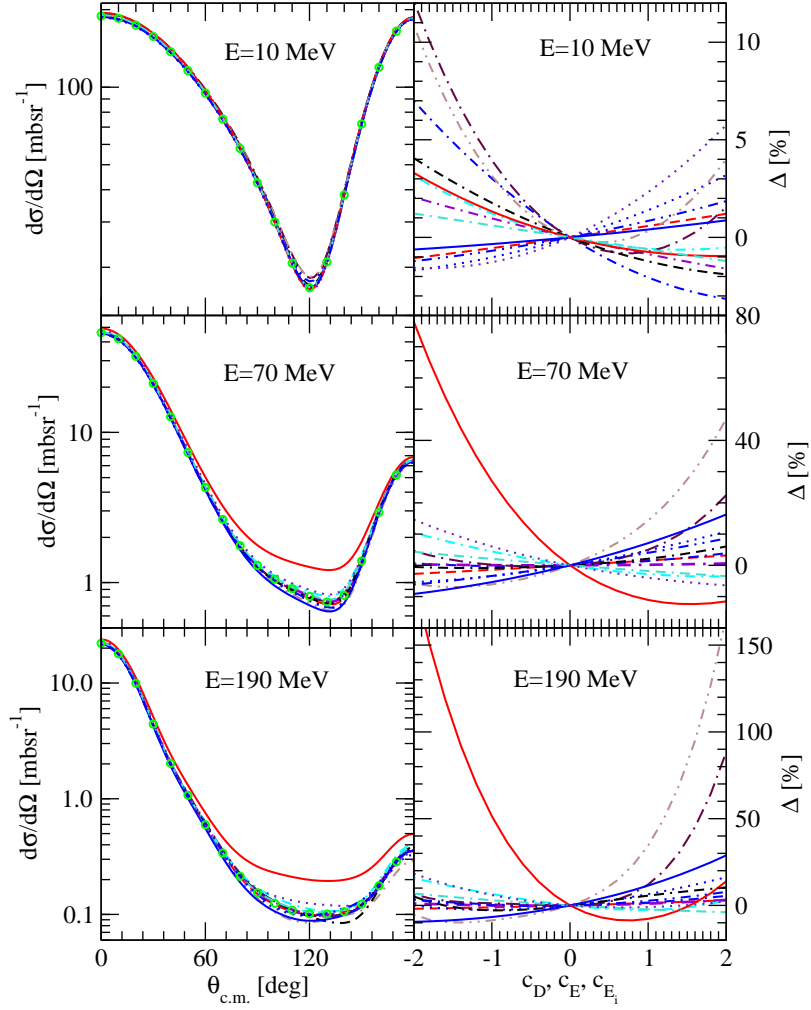}
\caption{
  (color online) (left column) The elastic nd scattering differential cross section
  $\frac {d\sigma} {d\Omega}$ at the incoming neutron laboratory energies
  $E=10, 70,$ and $190$~MeV. The lines depicted
  by (green) circles show the results obtained with the SMS N$^4$LO$^+$ NN 
  potential with the
  regularization parameter $\Lambda=450$~MeV, supplemented by the parameter-free
  2$\pi$-exchange N$^2$LO 3NF. Other lines are the  results when the above
  dynamics is augmented with a single contact term of strength
  $c_i=-1.0$: $D$ - (red) short-dashed,
  $E$ - (blue) short-dashed-dotted, $E_1$ - (blue) dotted,
  $E_2$ - (violet) short-dashed-dotted, $E_3$ - (cyan) short-dashed-dotted,
  $E_4$ - (maroon) long-dashed-dotted,
  $E_5$ - (brown) short-dashed-double-dotted,
  $E_6$ - (black) double-dashed-dotted,  $E_7$ - (blue) solid,
  $E_8$ - (red) solid, $E_9$ - (turquoise) double-dashed-dotted,
  $E_{10}$ - (indigo) dotted, and $E_{13}$ - (blue) dashed-double-dotted.
  In the right column a percentage deviations $\Delta$ (see text)
  of the  single  contact term  predictions
  with respect to the  parameter-free 
 part of the N$^2$LO 3NF  $V(\theta_0)$ are shown as a function
  of the strength $c_i$. The lines in the right column correspond to those in
  the left column.
}
\label{fig1}
\end{figure}

\begin{figure}    
\includegraphics[scale=0.7,clip=true]{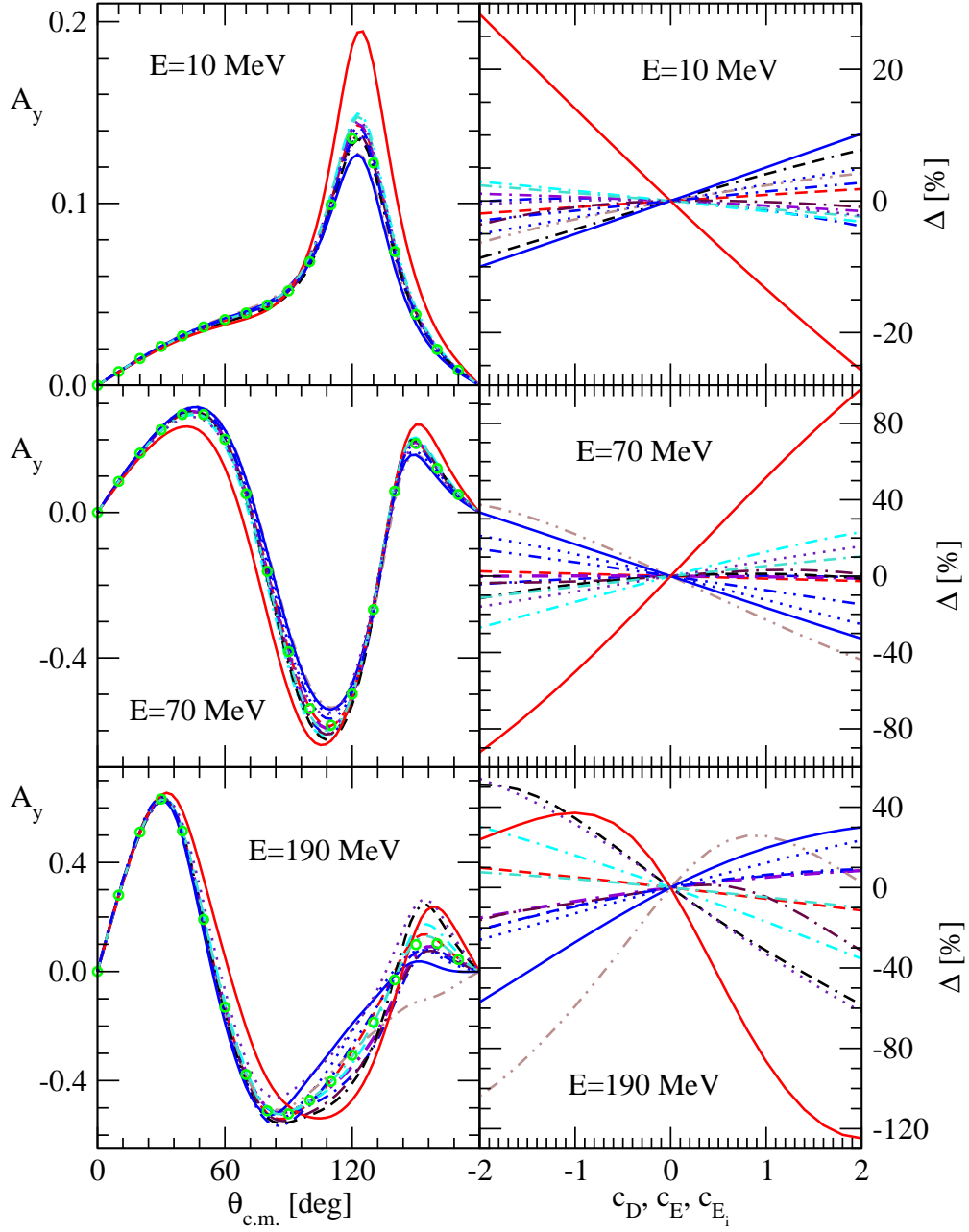}
\caption{ 
(color online) The same as in Fig.\ref{fig1} but for the 
   nucleon analyzing power $A_y$. 
}
\label{fig2}
\end{figure}

\begin{figure}    
\includegraphics[scale=0.7,clip=true]{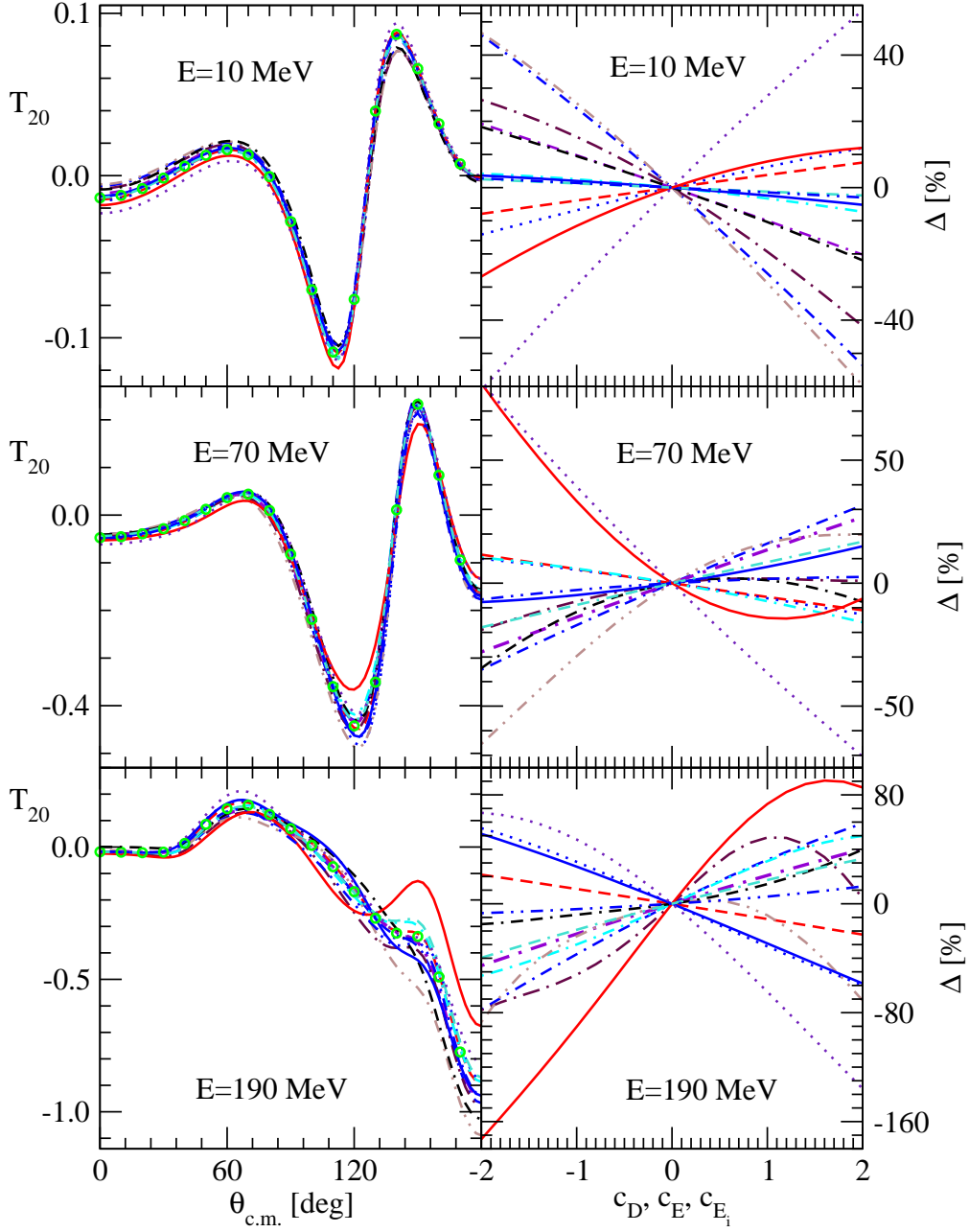}
\caption{
(color online) The same as in Fig.\ref{fig1} but for the 
   deuteron tensor analyzing power $T_{20}$. 
}
\label{fig3}
\end{figure}

In Figs.\ref{fig1}-\ref{fig3} we show the results 
of this investigation for three  observables:
the differential cross section $\frac {d\sigma} {d\Omega}$, the
nucleon analyzing
power $A_y$, and the deuteron tensor analyzing power $T_{20}$, at
three energies $E=10$, $70$, and $190$~MeV. In the left column predictions for
these observables, obtained with  parameter free part of
the N$^2$LO 3NF $V(\theta_0)$,  
as well as  including in addition consecutively
 each of the 13 short range terms, 
 are shown as a function of the c.m.  scattering angle $\theta$. 
 In the right column the quantity $\Delta$ is displayed
  for that particular observable
  and for   each of the 13 short range terms, as a function of their strengths.

  The pattern of sensitivities exemplified in Figs.\ref{fig1}-\ref{fig3}
  reflects the features common for all studied
  observables. At low energy $E=10$~MeV
  the changes induced by different short-range components are of the order
  of about few percent, with the exception of few terms, which 
  affect significantly 
  a particular observable (see for example 
  Fig.~\ref{fig2} displaying the dominating impact of the $E_8$ term on $A_y$). With
  increasing energy the pattern changes both with respect to the magnitudes
  of the induced changes and with respect to the number of appreciably contributing terms. For the
  cross section (see Fig.\ref{fig1}) the magnitude of $\Delta$ reaches
  approximately 100~\%
  at $190$~MeV with the dominating
  contributions coming from the $E_5$ and $E_8$ terms. Similarly, for $A_y$
  (Fig.\ref{fig2}) and $T_{20}$ (Fig.\ref{fig3}) $E_8$ prevails at higher 
  energies and its effects have opposite signs at $70$~MeV
  and $190$~MeV. Such alternating patterns of sensitivities 
  with respect to the observable, energy,
  and contributing short-range contact term, foreshadow a successful 
  determination of all the contact terms strengths
  by fitting theoretical predictions to Nd elastic scattering data. 

  It should be emphasised that practically for most of the studied
  observables and energies,
  the contributions from the N$^2$LO $D$ and $E$ contact terms,
  expressed in terms
  of $\Delta$, do not belong to the most significant ones. In view of that, to
  determine how important the N$^4$LO contact terms are for
  the 3N bound system, we
  calculated their expectation values in the triton arising from the 
   N$^4$LO$^+$ SMS NN force ($\Lambda=450$~MeV) alone,  
  taking their
  strengths $c_i=1.0$. (see Table~\ref{tab1}). It is clear that nearly all
  short-range terms, with the exception of the $E_7$ and $E_8$ terms providing
  negligible contributions, make
  comparable and significant contributions to the triton
  potential energy. Therefore the subleading
  contact terms seem to play a significant role not only in 3N continuum but
  also in the bound states.

\section{Fixing strengths of contact terms}
\label{results} 

Let us come back to the basic question, how precisely and reliably 
all the strengths of the 3NF short-range components can be determined. 
The results of the previous section are promising, since they 
reveal a pattern of high sensitivity (changing with the observable and energy) 
to practically all the short-range components of the considered chiral 3NF
but this is only a necessary condition. 
The decisive is the quality and number 
 of available Nd elastic scattering data points.
 As a consequence of strenuous efforts of experimentalists the
 amount and precision of the available pd data has recently radically improved. Still, the
 presently accessible pd
data base is not so numerous as the proton-proton (pp) one. 
Below the pion production threshold pd data have been taken at a
 number of energies on
the elastic scattering cross section, proton analyzing power,
all deuteron vector and tensor analyzing powers, and in few cases some
polarization transfer and spin correlation coefficients. For the nd data basis
the situation is worse and only at few energies nd elastic scattering
cross sections and neutron analyzing powers are available, with larger
errors than in the case of the pd data. In addition, for the nd system
also high precision data for the total nd cross section have been collected.

 In the first step we investigate if  
 the alternating pattern of large sensitivities found in the previous section
and access to high quality Nd data 
 would be sufficient for a successful determination of all the strengths. 
 To be specific, we assume that we have high quality data for the
 cross section, the nucleon analyzing power and
 the deuteron vector and tensor analyzing
powers. We generated such pseudo-data at five energies $E=10, 70, 135, 190,$
and $250$~MeV, 
using our dynamical input and taking all 13 strengths $c_i=1.0$. The data
covered the range of the c.m. angles $\theta_{c.m.} \in (40^{\circ} - 170^{\circ})$ with a 
step of $5^{\circ}$ and had the preconditioned relative error of $5~\%$. 
To these data we applied the least-squares method by introducing the $\chi^2(c_i)$
merit function:
\begin{eqnarray}
 \chi^2(c_i) &=& \sum_{Obs,\theta_k,E} [  \frac {Obs^{th}(c_j,\theta_k,E)
    -Obs^{ex}(\theta_k,E)}     {\Delta Obs^{ex}(\theta_k,E)} ]^2~,
\label{eq20}
\end{eqnarray}
and looked for minimum of $\chi^2(c_i)$ with respect to the strengths $c_i$.
To find the minimum we applied
 the Levenberg-Marquard method \cite{num-rec,marquardt},
which, in addition to $\chi^2$ values, requires also the gradient of $\chi^2$
with respect to the parameters $c_i$. Since the dependence of the
elastic scattering transition amplitude U (Eq.(\ref{eq3})) on $c_i$ has
the form:
\begin{eqnarray}
 U &=& \bar U + \sum_i c_i U_i + \sum_{i,k} c_i c_k U_{ik} ~, 
\label{eq21}
\end{eqnarray}  
the gradient of $\chi^2$ is quickly accessible for any set of
strengths $c_i$. 

Starting from different sets of initial values of
strengths $c_i$ we found that it is relatively easy to reproduce very
accurately the values of strengths incorporated in the pseudo-data.
We succeeded in all
cases to reproduce
the input strengths by fitting observables at individual energies as well as
performing a multi-energy search when including all energies.

\begin{figure}
\begin{tabular}{c}
\resizebox{140mm}{!}{\includegraphics[scale=0.6,clip=true]{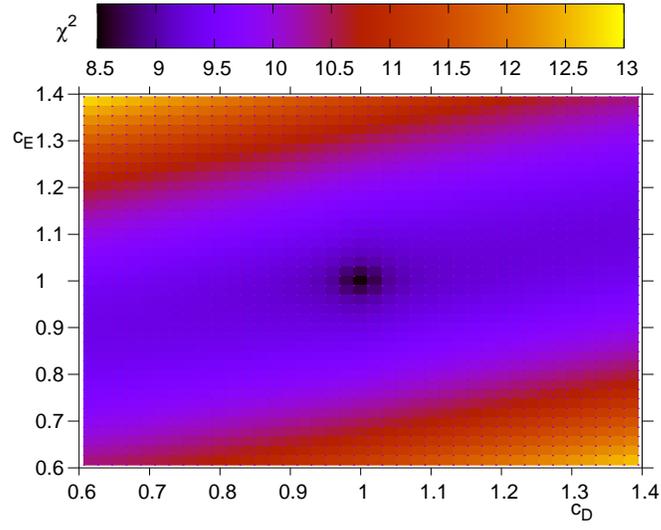}} \\
\resizebox{140mm}{!}{\includegraphics[scale=0.6,clip=true]{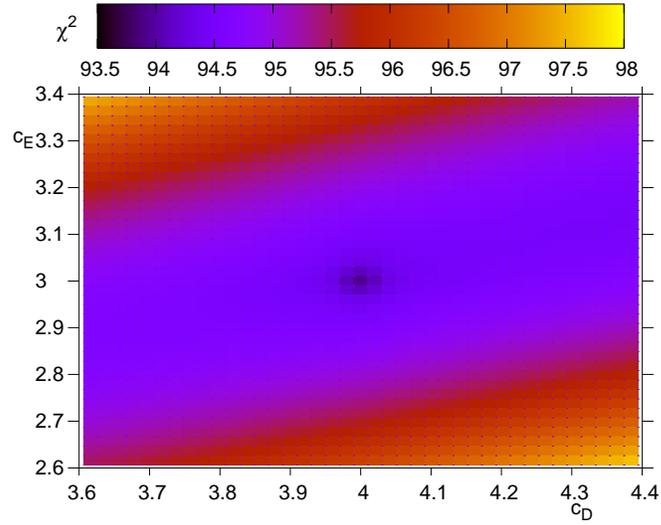}} \\
\end{tabular}
\caption{
(color online) 
  The maps of $\chi^2$ per datum point ($\chi^2/N$) values  from fitting
  the pseudo-data at $E=70$~MeV with complete 
  (up) and incomplete (down) theory (see text for explanations).
}
\label{fig4}
\end{figure}

\begin{figure}    
\includegraphics[scale=0.6,clip=true]{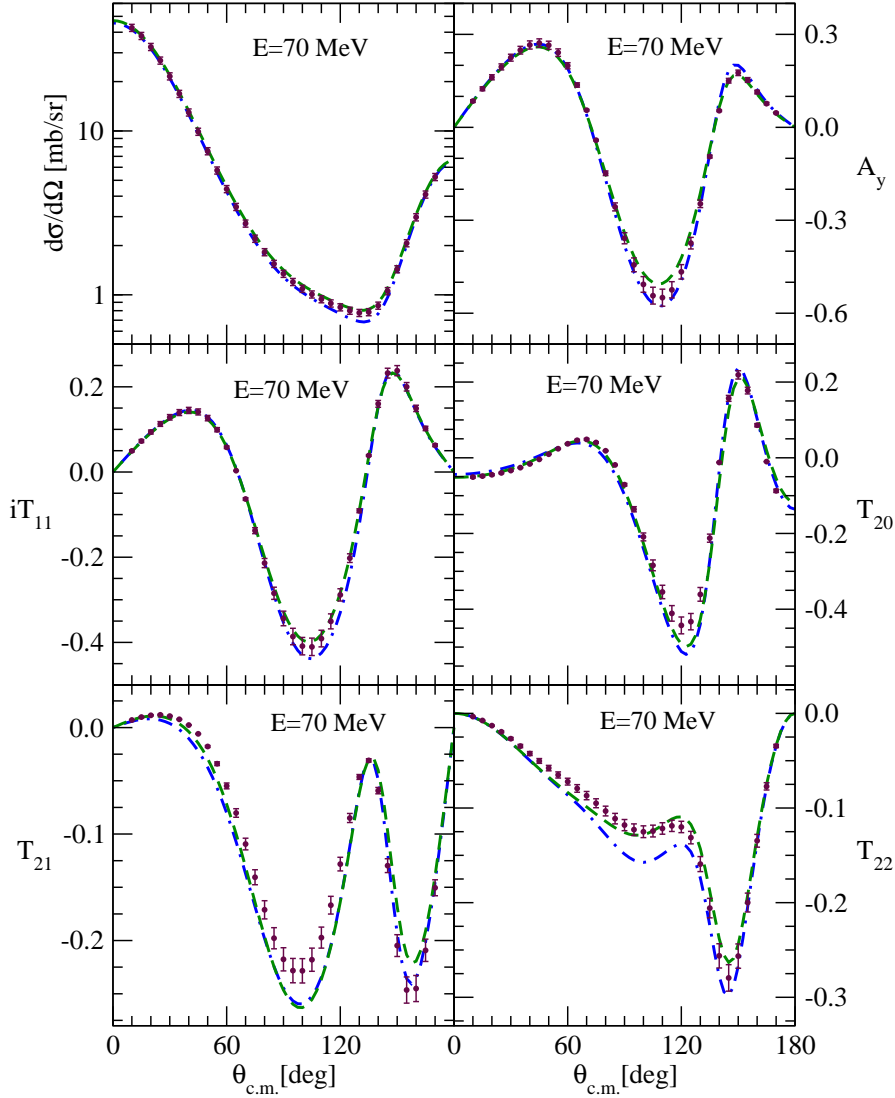}
\caption{
  (color online) (left column) The influence of the lacking dynamics on the results
  of the least-squares fit on the example of pseudo-data ((maroon) circles)
  for elastic nd scattering
  differential cross section
  $\frac {d\sigma} {d\Omega}$ as well as for all the nucleon and deuteron
  analyzing powers: $A_y$, $iT_{11}$, $T_{20}$, $T_{21}$, and $T_{22}$,
  at $E=70$~MeV. The pseudo-data were generated with our emulator
  using the SMS N$^4$LO$^+$ NN  potential with the
  regularization parameter $\Lambda=450$~MeV, supplemented with the N$^2$LO 3NF
  with the strengths of contact terms $c_D=1.0$ and $c_E=1.0$.
  To each data point 
	a relative error of $5\%$
	was prescribed. The (blue) dashed-dotted line is the
  result of 3N calculations  with the above-defined  dynamics but
  omitting the parameter free
  2$\pi$-exchange N$^2$LO 3NF term. The (green) dashed line is the result
  of the least-squares fit to the pseudo-data with this lacking dynamics,
  which provided
  values of $c_D=3.98 \pm 0.08$ and $c_E=2.99 \pm 0.03$.
  }
\label{fig5}
\end{figure}

When studying that point we took a step further and investigated the situation in
which the data are fitted with an incomplete theory, that is 
when some parts of the underlying dynamics are missing. Evidently such a situation occurs
when we try to fix strengths of short-range contact terms without the N$^3$LO
3NF components included in the calculations. To study such a case
we generated
similarly as previously pseudo-data at $E=70$~MeV taking chiral dynamics
based on N$^4$LO$^+$ NN SMS interaction supplemented by the complete
 N$^2$LO 3NF with
strengths of the D and E terms $c_D=c_E=1.0$. To such pseudo-data we performed
a least-squares fit  in order to fix strengths $c_D$ and $c_E$,
using original
dynamics, omitting however parameter-free $2\pi$-exchange term in the N$^2$LO
3NF. The  results are shown in Fig.\ref{fig4} in the form of maps of
$\chi^2$ per datum point values in space of $c_D-c_E$. With complete
dynamics we recovered easily the incorporated strengths and obtained
$c_D=1.00 \pm 0.08$ and $c_E=1.00 \pm 0.03$, reaching at that point values of
$\chi^2 = 0.0 $. When the fit was performed with incomplete dynamics we found a
significant shift of the $\chi^2$ minimum position to larger values of
$c_D=3.98 \pm 0.08$ and $c_E=2.99 \pm 0.03$, with concurrent deterioration of
the quality of data description as evidenced by increased minimal
 values of $\chi^2$ per datum point $\chi^2/N \approx 80$ compared to
$\chi^2/N \approx 0$ for complete dynamics.
In Fig.\ref{fig5} we present in more detail the quality of
pseudo-data description
by showing pseudo-data themselves ((maroon) circles) and
predictions obtained with the
incomplete dynamics ((blue) dashed-dotted lines). The (green) dashed line
is the result of the least-squares fit to the pseudo-data with
incomplete dynamics,
which, in general,  improves slightly description of
data but at the expense of increased
strengths of short-range terms D and E.

Since the essential element, namely the N$^3$LO components of the chiral 3NF, is
missing in our dynamics, in view of above  it is evident that results and
conclusions of the present investigation have to be treated with caution
and this study must be considered as preliminary. It should be repeated
when the N$^3$LO 3NF components become available. Nonetheless, having
that restriction in mind, we are ready to answer  our main question.
To this end we prepared Nd elastic scattering data
base at the five energies $E=10, 70, 135, 190,$ and $250$~MeV, collecting
data points 
for the differential cross section, the nucleon vector analysing power, and the
deuteron vector and tensor analysing powers, which reflects more or less the
status of the presently available Nd data and which are listed
in Table~\ref{tab2}, and performed multi-energy least-squares fit to data at
three energies ($E=10, 70,$ and $135$~MeV). Since our
3N continuum calculations neglect the proton-proton (pp) Coulomb force,
whose effects in elastic pd scattering are restricted mostly to
small energies and
forward c.m. angles, we took only pd data at $\theta_{c.m.} > 40^{\circ}$
when calculating $\chi^2$ (altogether $786$ data points). 

The resulting values of strengths $c_i$ 
are listed in Table~\ref{tab3} together with errors (standard deviations)
 obtained from the
covariance matrix $C(c_i,c_j)$ shown in Table~\ref{tab4}.
The four strengths which have large magnitudes belong
to the subleading order N$^4$LO: $c_{E_1}=6.40$,
$c_{E_2}=7.80$, $c_{E_3}=6.97$, and $c_{E_7}=-7.40$. 
It is interesting to note that only strengths in this order, mostly those with
large magnitude, $c_{E_1}$, $c_{E_2}$, and $c_{E_3}$, are
 strongly correlated as evidenced by the 
values of the corresponding correlation coefficients:
$\rho(c_{E_1},c_{E_2})=0.95$, $\rho(c_{E_1},c_{E_3})=0.98$, 
$\rho(c_{E_2},c_{E_3})=0.93$. The strength $c_{E_3}$ is also strongly
correlated with $c_{E_4}$:
$\rho(c_{E_3},c_{E_4})=-0.92$, and  $c_{E_7}$  with $c_{E_8}$:
$\rho(c_{E_7},c_{E_8})=0.99$. 
  There is
  only a small correlation between $c_D$ or $c_E$  and all subleading
  terms as well as between $c_D$ and $c_E$ themselves. The final
  value of $\chi^2$ per data point $\chi^2/N \approx 35$ indicates that the
  quality of data description is notably inferior than in the case of the 
  nucleon-nucleon system. 
  The large value of final $\chi^2/N$  as well as large magnitudes
  of some strengths probably reflect the omission of
  the N$^3$LO term in the 3NF. Therefore the present investigation should be
 repeated when this term is available.

\begin{figure}    
\includegraphics[scale=0.56,clip=true]{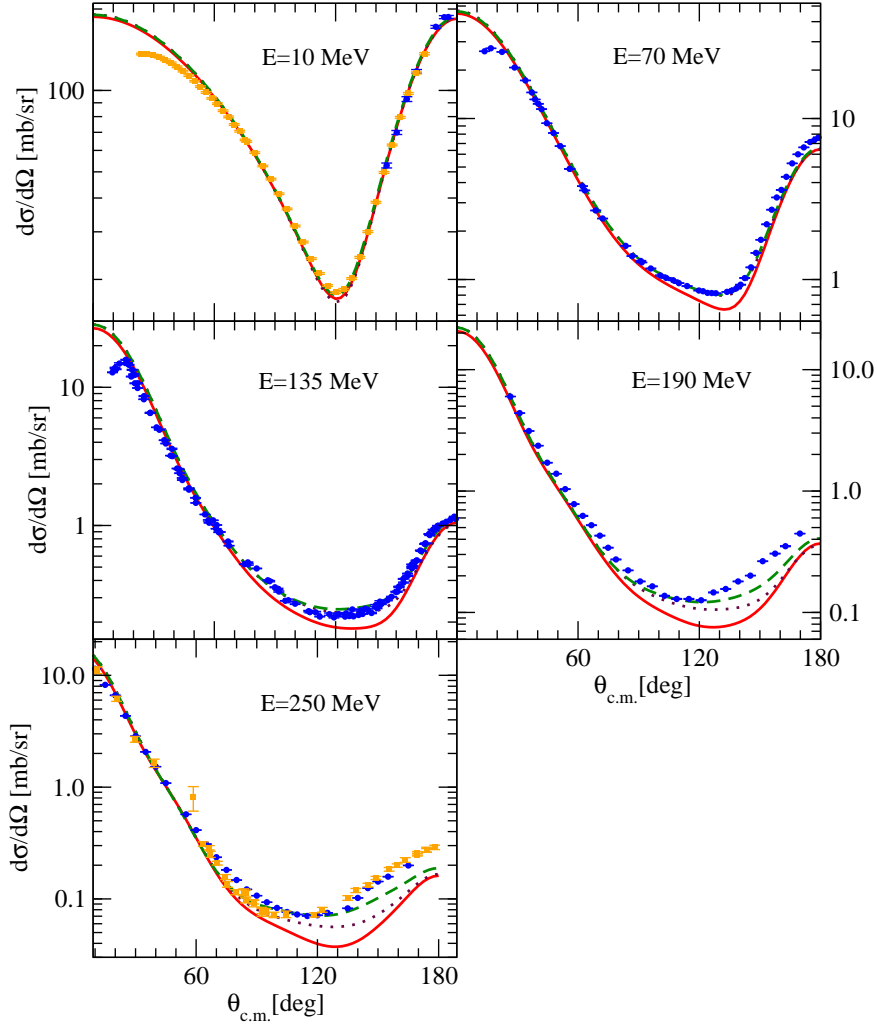}
\caption{
  (color online) The elastic Nd scattering differential cross section
  $\frac {d\sigma} {d\Omega}$ at the incoming nucleon laboratory energies
  $E=10, 70, 135, 190,$ and $250$~MeV. The (red) solid lines 
   were obtained with the SMS N$^4$LO$^+$ NN   potential with the
  regularization parameter $\Lambda=450$~MeV. When that potential is
  supplemented with the N$^2$LO 3NF with the strengths of the contact terms
  $c_d=2.0$ and $c_E=0.2866$  (combination reproducing the $^3$H binding energy
  and providing a good description of the $70$~MeV pd cross sections) 
  predictions are displyed with the (maroon) dotted lines.
  The (green) dashed lines show the results obtained with
   the strengths of contact terms presented in Table~\ref{tab3}, fixed in 
  the multi-energy least-squares fit to data at $E=10, 70,$ and $135$~MeV
  (shown in Table~\ref{tab2}). 
  The (blue) circles and (orange) squares
  are $10$~MeV nd data from \cite{how_sig10} and pd data from \cite{sagara}, 
  respectively, The  (blue) circles at other energies are pd data from:
  $70$~MeV \cite{seki140}, $135$~MeV \cite{seki140,sak99},
  $190$~MeV \cite{ermisch}, $250$~MeV \cite{hatanaka}.
   The (orange) squares at $250$~MeV are $248$~MeV nd data of \cite{maeda250}.
}
\label{fig6}
\end{figure}

\begin{figure}    
\includegraphics[scale=0.6,clip=true]{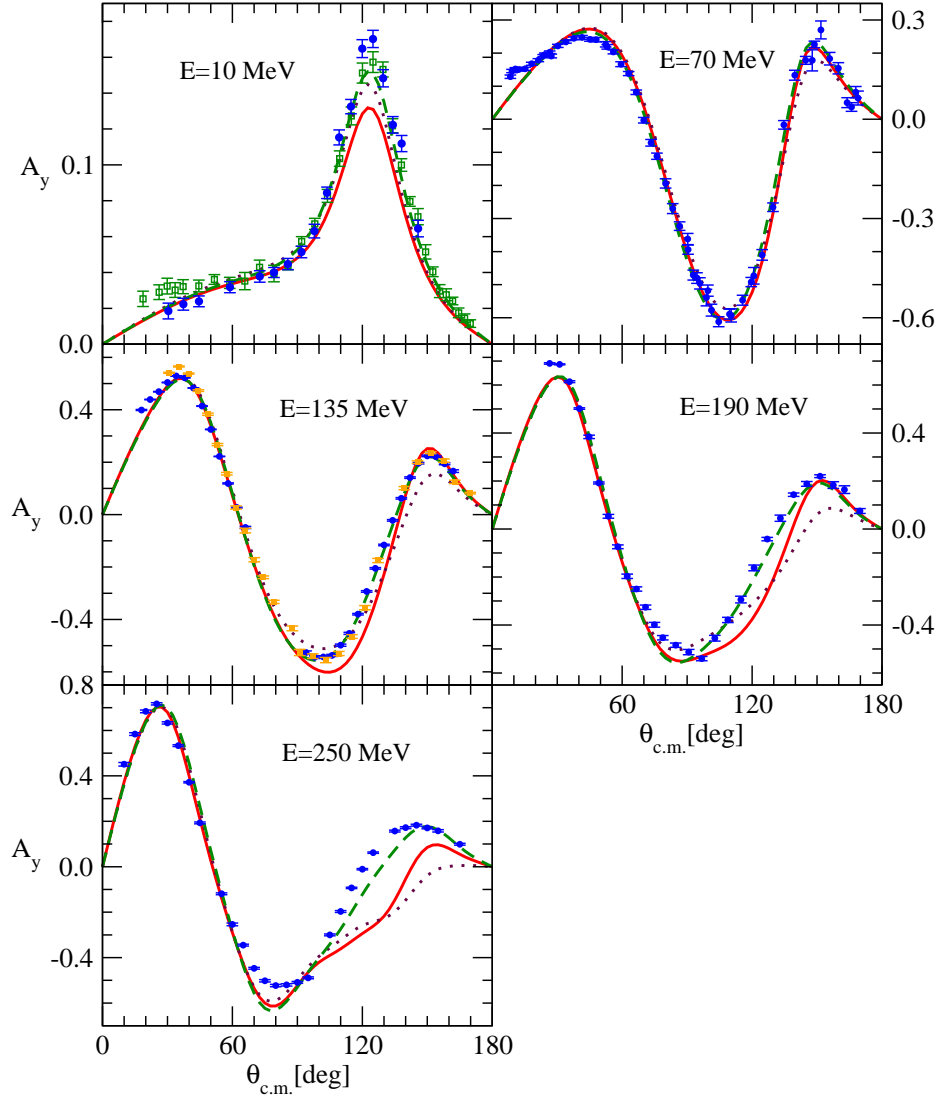}
\caption{
  (color online) The same as in Fig.\ref{fig6} but for the
  nucleon analyzing power $A_y$. The data are from: $10$~MeV (blue) circles nd
  data \cite{how_ay10} and (green) squares pd data \cite{sperisen84},
  $70$~MeV (blue) circles pd data (at $65$~MeV) \cite{shimizu}, 
   $135$~MeV (blue) circles pd data \cite{iucf2006}
  (orange) squares pd data \cite{ermisch}, $190$~MeV (blue) circles
  pd data \cite{ermisch},  $250$~MeV (blue) circles pd data \cite{hatanaka}.
}
\label{fig7}
\end{figure}

\begin{figure}    
\includegraphics[scale=0.6,clip=true]{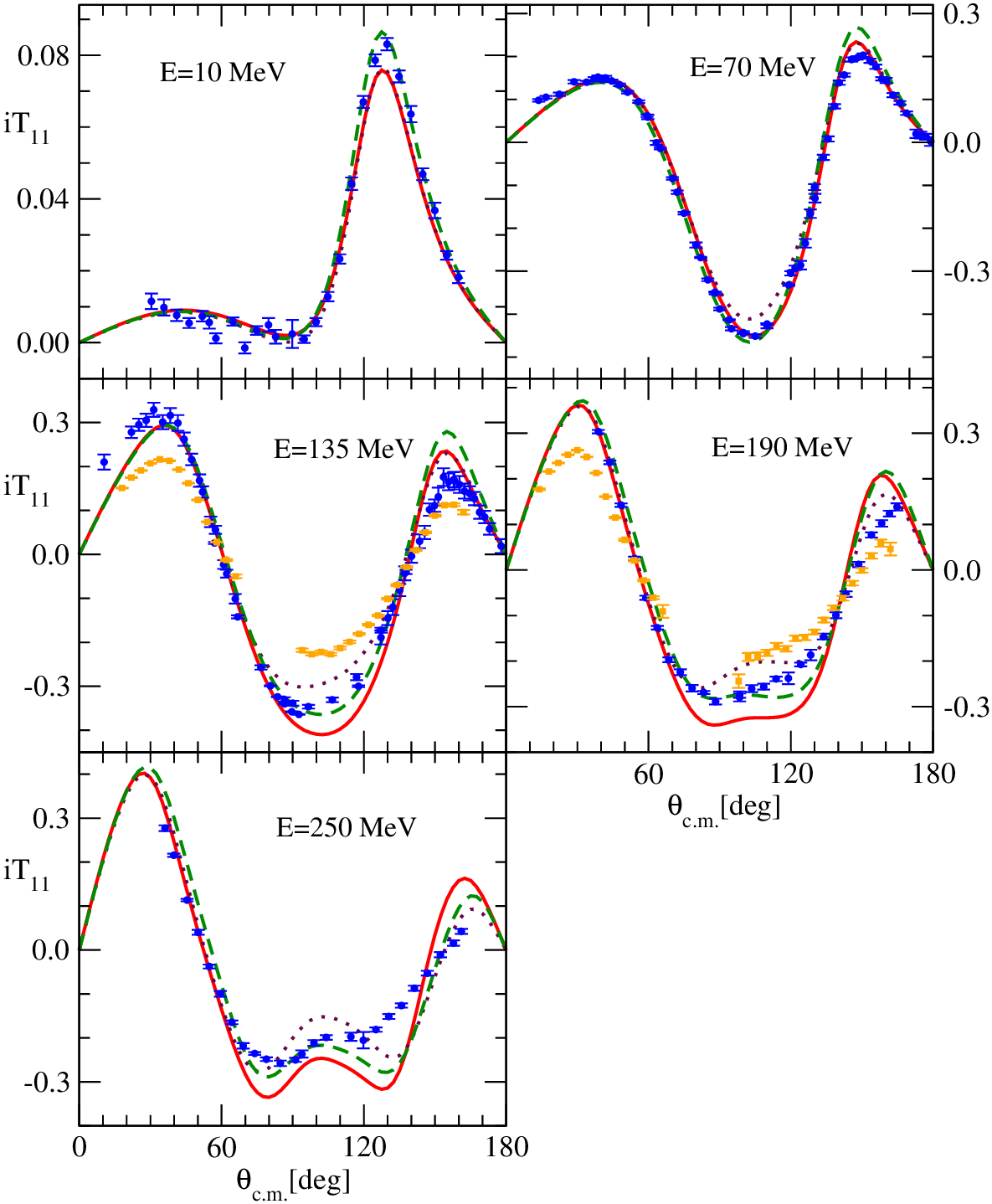}
\caption{
  (color online) The same as in Fig.\ref{fig6} but for the
  deuteron vector analyzing power $iT_{11}$.
  The data are from: $10$~MeV (blue) circles pd data \cite{sawada83,sperisen84},
  $70$~MeV (blue) circles  pd data
\cite{seki140},  $135$~MeV (blue) circles pd data \cite{iucf2006}
  (orange) squares pd data \cite{ermisch}, $190$~MeV (blue) circles
pd data \cite{seki190} (orange) squares pd data (at $200$~MeV) \cite{iucf2006},
$250$~MeV (blue) circles pd data \cite{hatanaka}.
}
\label{fig8}
\end{figure}

\begin{figure}    
\includegraphics[scale=0.6,clip=true]{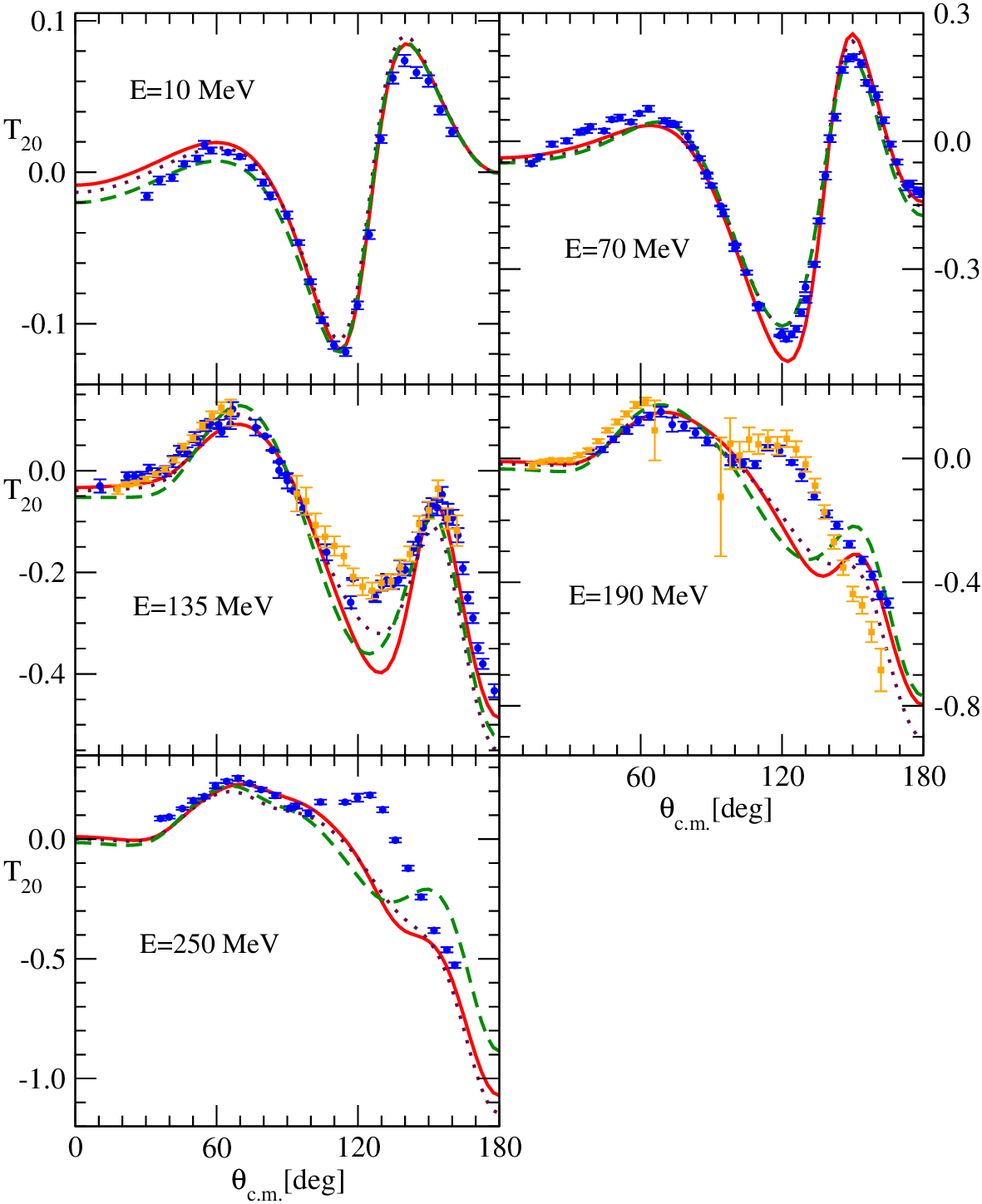}
\caption{
  (color online) The same as in Fig.\ref{fig6} but for the
  deuteron tensor analyzing power $T_{20}$.
  The data are from: $10$~MeV (blue) circles pd data \cite{sperisen84},
  $70$~MeV (blue) circles pd data \cite{seki140},
  $135$~MeV (blue) circles pd data \cite{seki190}
  (orange) squares pd data \cite{iucf2006}, $190$~MeV (blue) circles
pd data \cite{seki190} (orange) squares pd data (at $200$~MeV)
 \cite{iucf2006},  $250$~MeV (blue) circles pd data \cite{seki250}.
}
\label{fig9}
\end{figure}

\begin{figure}    
\includegraphics[scale=0.6,clip=true]{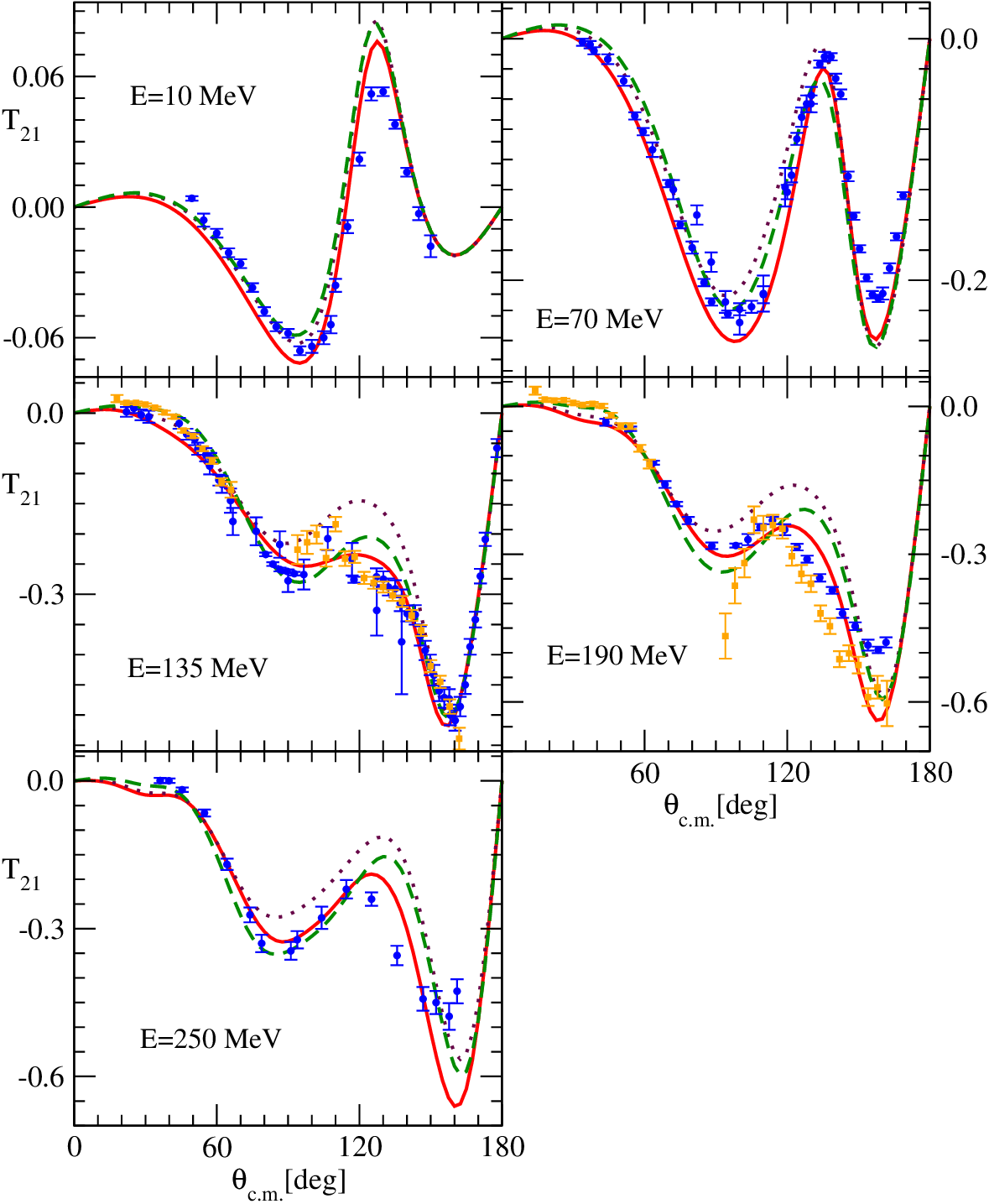}
\caption{
  (color online) The same as in Fig.\ref{fig6} but for the
  deuteron tensor analyzing power $T_{21}$.
  The data are from: $10$~MeV (blue) circles pd data \cite{sperisen84},
  $70$~MeV (blue) circles pd data \cite{seki140},  $135$~MeV (blue) circles
  pd data \cite{seki190}
  (orange) squares pd data \cite{iucf2006}, $190$~MeV (blue) circles
  pd data \cite{seki190} (orange) squares pd data (at $200$~MeV)
  \cite{iucf2006},
  $250$~MeV (blue) circles pd data \cite{seki250}.
}
\label{fig10}
\end{figure}

\begin{figure}    
\includegraphics[scale=0.6,clip=true]{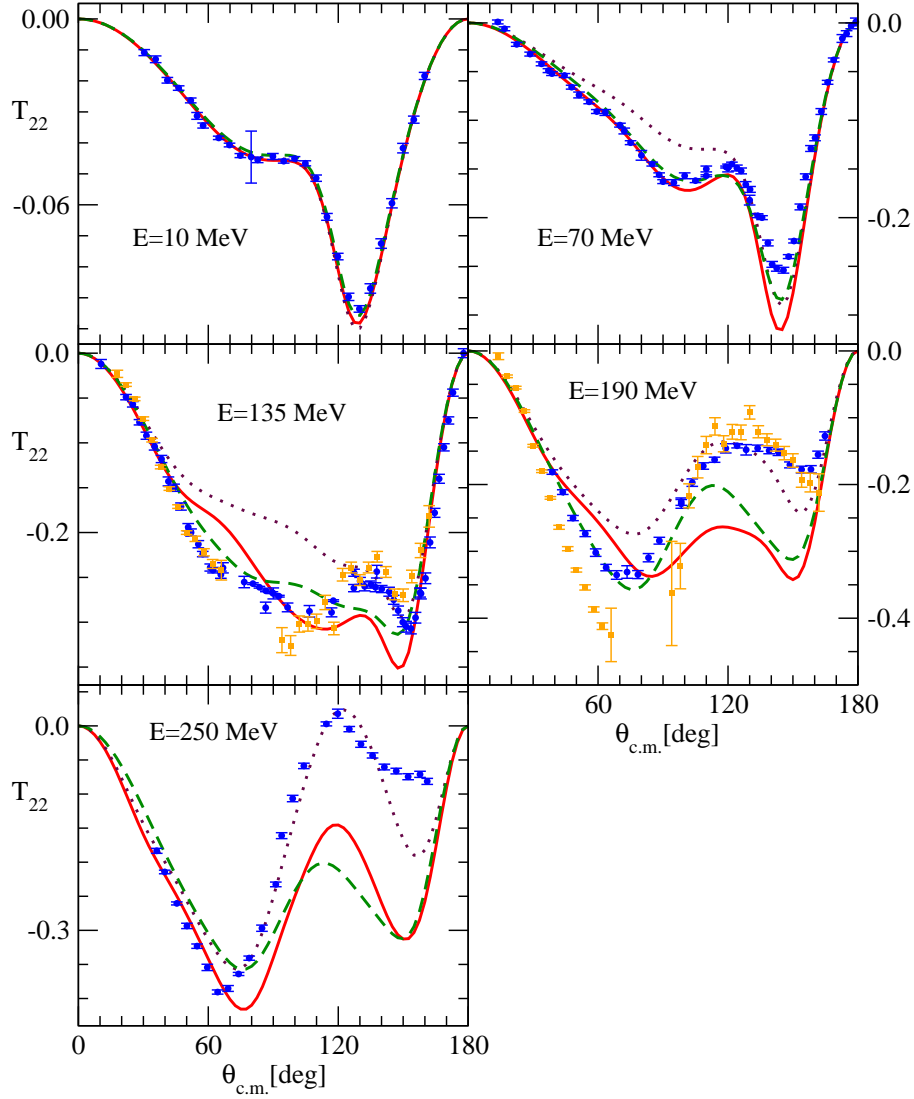}
\caption{
  (color online) The same as in Fig.\ref{fig6} but for the
  deuteron tensor analyzing power $T_{22}$.
  The data are from: $10$~MeV (blue) circles pd data \cite{sperisen84},
  $70$~MeV (blue) circles  pd data \cite{seki140},
  $135$~MeV (blue) circles pd data \cite{seki140}
  (orange) squares pd data \cite{iucf2006}, $190$~MeV (blue) circles
  pd data \cite{seki190} (orange) squares pd data (at $200$~MeV)
  \cite{iucf2006},
  $250$~MeV (blue) circles pd data \cite{seki250}.
}
\label{fig11}
\end{figure}

 In Figs.\ref{fig6}-\ref{fig11} we show how well the data from our basis 
   (the (green) dashed lines) are described
 by the 3N Hamiltonian with fixed in this way strengths of 
 contact terms.  Since the least-squares fit was
performed for data at 3 lowest energies the results at $190$ and $250$~MeV
should be considered as predictions. To asses the magnitudes of the contact terms' effects
we show also predictions based on the NN SMS N$^4$LO$^+$ potential
(the (red) solid lines) and the results obtained when the latter was augmented by the N$^2$LO 3NF 
with the strengths of D and E terms, $c_D=2.0$, $c_E=0.2866$, determined from the $^3$H binding
energy and the $70$~MeV pd cross section (the (maroon) dotted lines).

In nearly all cases the fit to data improves significantly the description of
not only fitted data but also the data at the two largest energies. It is very
clear, especially for the cross section (see Fig.\ref{fig6}),
where the discrepancy between data and
theory, found in the region of the cross section minimum up to the backward c.m.
angles, is practically removed at $70$ and $135$~MeV. At $190$ and $250$~MeV
 the inclusion of N$^4$LO contact terms  brings the theory closer to data.

For the nucleon $A_y$ and the deuteron vector $iT_{11}$ analyzing powers 
there is a significant improvement of
the data description in the maximum of the analyzing power at $10$~MeV
 (see Figs.\ref{fig7} and \ref{fig8}). That effect was also
found below the deuteron breakup threshold in Ref.~\cite{girlanda2019} and
supports the conclusion of ~\cite{girlanda2019} that the low energy analyzing
power puzzle may probably find its solution in the subleading N$^4$LO  3NF contact terms.  

A similar picture emerges for the tensor analysing powers
(see Figs.\ref{fig9}-\ref{fig11}); here, however, at the largest energies big
discrepancies to data remain.

\begin{figure}    
\includegraphics[scale=0.6,clip=true]{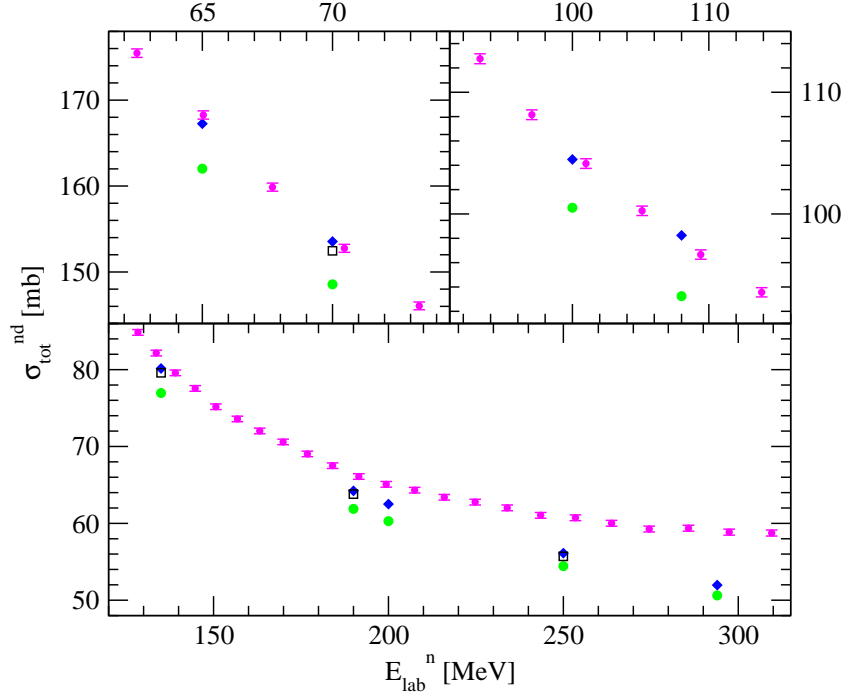}
\caption{
  (color online) The total cross section for neutron-deuteron scattering.
  The (green) circles are predictions of the  SMS N$^4$LO$^+$ NN 
  potential with the regularization parameter $\Lambda=450$~MeV. That
  potential supplemented by N$^2$LO 3NF
  with strengths $c_D=2.0$ and $c_E=0.2866$
  gives (blue) diamonds (combination of strengths reproducing
  the $^3$H binding energy and the $70$~MeV pd cross section data).
  The (green) squares are the total nd cross sections obtained with the strengths of
	N$^2$LO and N$^4$LO contact terms fixed in the multi-energy ($E=10, 70,$ and
	$135$~MeV)  least-squares fit to Nd data,
  shown in Table~\ref{tab3}.
  The (magenta) circles are the nd data from \cite{abfalt98}.
}
\label{fig12}
\end{figure}

The large advancement in the description of the elastic Nd scattering
cross section
documented 
in Fig.\ref{fig6} at the two largest energies prompted us to verify
the situation for the total nd scattering cross section.
In Fig.\ref{fig12} we show at a few energies the
SMS N$^4$LO$^+$ NN potential predictions (the (green) circles) together with
results calculated with this NN force combined 
with the N$^2$LO 3NF ((blue) diamonds). We display also the total
cross section data
from Ref.~\cite{abfalt98} ((magenta) circles). Additionally the
total cross sections
 obtained with the contact terms fixed by the least-squares fit
 ((green) squares) are shown at the selected four energies.
 Up to $135$~MeV the inclusion of the 3NF
(N$^2$LO or N$^2$LO combined with contact N$^4$LO terms) agrees with the total
cross section data. However, at $190$ and $250$~MeV even the addition of
N$^4$LO contact terms, what significantly improved the description of the elastic
Nd scattering cross section, does not help to remove the growing with energy gap
between data and theory. It means that very likely
 3NF is not responsible for that discrepancy. 
Since at these energies pion production starts to play a role, it is very
probably that this new channel, not taken into account in our purely nucleonic
scheme, is responsible for this discrepancy.

\begin{figure}    
\includegraphics[scale=0.6,clip=true]{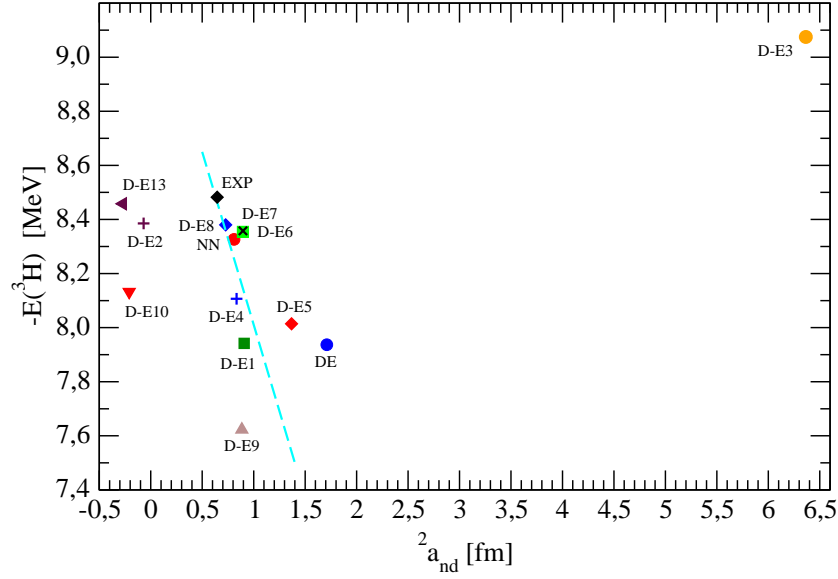}
\caption{
  (color online) The doublet nd scattering length $^2a_{nd}$
  and the triton binding energy $E_{^3H}$
  (in 18 channel calculations ($j_{max}=2$)) for different combinations of
  2N and/or  3N forces. The (red) circle is the result
   for the  SMS N$^4$LO$^+$ NN 
  potential with the  regularization parameter $\Lambda=450$~MeV.
  Other symbols show results for that NN potential combined with
  the N$^2$LO 3NF and supplemented by a sum of the consecutive N$^4$LO contact terms
  (all contact terms with strengths from Table~\ref{tab3}:
  (blue) circle (DE) - D+E N$^2LO$, (green) square (D-E1=D+E+E1),
  (maroon) plus (D-E2=D+E+E1+E2), (blue) plus (D-E4),
  (red) diamond (D-E5), (green) square (D-E6), 
  (black) x (D-E7), (blue) diamond (D-E8), (brown) triangle up (D-E9),
  (red) triangle down (D-E10), (maroon) triangle left (D-E13). The (cyan)
  dashed line is a Phillips line for (semi)phenomenological interactions from
  Ref.~\cite{philips}. The (black) diamond shows the experimental values of
  $^2a_{nd}=0.645 \pm 0.003$~fm ~\cite{schoen} and $E_{^3H}=-8.4820(1)$~MeV.
}
\label{fig13}
\end{figure}

In investigations of the 3N continuum  performed up to now with chiral forces,
only N$^2$LO components of a 3NF were included and
the experimental triton binding energy
  was essential for determining the low energy constants $c_D$ and
  $c_E$, by providing a set of pairs ($c_D$,$c_E$) which reproduced that basic
  quantity (forming the so-called ``correlation line''
  \cite{epel2019,lenpic2021,epel_tower}).
  In this way it was ensured that the triton energy is
  correctly reproduced.
  Since the doublet nd scattering length $^2a_{nd}$ is strongly correlated
  with the triton binding energy $E_{^3H}$  (often displayed in the form 
  of the so-called Phillips line \cite{philips}),
  it assures also a more or less correct description of this quantity.
In Fig.\ref{fig13} we show predictions for the triton binding energy and
  for the doublet scattering length $^2a_{nd}$ obtained with
  the 3N Hamiltonian based  on the SMS N$^4$LO$^+$ 
  NN potential combined with N$^2$LO 3NF together with
  all the subleading N$^4$LO contact terms with the values of strengths of the
  short-range components from
  Table~\ref{tab3} ((D-E13 maroon) triangle left). We show also 
  results for $E_{^3H}$ and $^2a_{nd}$  obtained by consecutive addition,
  to 2$\pi$-exchange term,  of
  contact terms,  starting from N$^2$LO 3NF (only $D$ and $E$ terms added:
  $DE$) and
  terminating when  all the N$^4$LO contact terms are
  added ($D+E+E_1+ \dots E_{13}$: $DE_{13}$). We display also the result
  for the SMS N$^4$LO$^+$ NN potential ((NN red) circle) and the 
   Phillips line ~\cite{philips}, along which predictions
  for $E_{^3H}$ and $^2a_{nd}$ of
  (semi)phenomenological NN potentials, alone or combined with standard
  3NF's, congregate. We observe large scattering of predictions around
  Phillips line for different combinations of the contact terms.
  Especially large
  deviation from the Phillips line occurs when contact terms up to $E_3$
  are added,
  leading to $E_{^3H}$ and $^2a_{nd}$ which are far away from the experimental
  values ($E^{exp}_{^3H}=-8.4820(1)$~MeV and $^2a^{exp}_{nd}=0.645 \pm 0.03 $~fm
  ~\cite{schoen}).
  It is evident that our 3N Hamiltonian is not able to reproduce
  the experimental values
  of the triton binding energy and the doublet $^2a_{nd}$ scattering length.  
     It seems that the strategy for
  determining the low energy constants applied when only N$^2$LO 3NF is 
  included in 3N calculations, needs to be modified when N$^4$LO short-range
  terms are also present. One has to forgo the
  correlation line and  incorporate the experimental triton binding energy
   in the fitting procedure in a different way. One possibility would be
  to include $^2a_{nd}$ in the fit using the same approach as for the 
  scattering, what by means of the Phillips line would probably provide
  the correct binding
  energy of $^3$H. Of course, with the availability of the N$^3$LO 3NF part
  it should be checked how far the description of $E_{^3H}$ and $^2a_{nd}$
  will be changed by a new set of determined strengths.

\section{Summary and conclusions}
\label{summary}

In this paper we investigate  the significance of the chiral 3NF 
contact terms for the description of the Nd elastic scattering observables 
for the incoming nucleon energies  up to the pion production threshold.
We used  the high precision SMS N$^4$LO$^+$ NN potential of Ref.~\cite{preinert}
 in combination with the N$^2$LO chiral 3NF supplemented by
 all the N$^4$LO contact terms. Our aim was to verify if it would be possible
 to fix strengths of all the contact terms by performing
 the least-squares fit of theory to Nd elastic scattering data.
 
The main results are summarized as follows.

\begin{itemize}

\item[-] In addition to the two contact terms of the N$^2$LO 3NF there are
  thirteen contact terms in the N$^4$LO 3NF, with two pairs
  being fully equivalent.
  Therefore a 3N Hamiltonian depends altogether on 13  parameters which
  are the strengths of those contact terms. They have to be found by fitting 
  theoretical predictions to 3N data. We found out that the pattern of
  sensitivities for  elastic Nd scattering observables to these 3NF components
  is diversified and changes with energy, observable and contact
  terms themselves. This provides a good base to fix the strengths of
  all the contact terms by
  fitting theoretical predictions to Nd data. It should be emphasised
  that even at lower energies the N$^2$LO contact terms are not the
  most essential and all the short range terms contribute equally.

\item[-] Using pseudo-data for the cross section and for a
  complete set of nucleon and deuteron analyzing powers, generated with our emulator, we
  checked that indeed it is possible to extract with high precision strengths
  of all the contact terms by the least-squares fit of
  theoretical predictions to such pseudo-data.
  Restricting to N$^2$LO 3NF only and neglecting parameter-free
  $2\pi$-exchange term in the 3NF we discovered implications
  of the missing dynamics on the results of such a procedure. There is a significant
  shift of determined strengths with concurrent deterioration of the quality of
  data description.

\item[-] Taking available Nd data for the cross section and a complete set
  of nucleon and deuteron analyzing powers at $10, 70,$ and $135$~MeV, we
  fixed strengths of all the contact terms by performing a least-squares fit
  to these data. With a 3N Hamiltonian defined in this way we received  not only a more
  satisfactory description of the fitted data but for
  most cases also  an improved description of the data taken at $190$ and $250$~MeV.
  Among others, we found a significant improvement of the description
  for the low-energy  analyzing power $A_y$ and $iT_{11}$ as well as for
   the cross section at higher 
   energies in the region around its minimum up to the backward angles.
     However, the large gap between the theory and data for the
  total nd cross section
  at energies above $\approx 200$~MeV remains, showing that it is
  not due to missing components of
  3NF but results probably from opening a new channel with real pion production.

\item[-] We found that the improved description of Nd elastic scattering data
  does not lead simultaneously to a good description of the triton binding energy
  and the doublet nd
  scattering length $^2a_{nd}$. Especially the scattering length
  obtained with fixed strengths of the contact terms lies far away
  from its experimental value. It will
  be interesting to see if inclusion of N$^3$LO 3NF component will improve 
  the description of these two quantities. 

\end{itemize}
  
It should be stressed out that in our dynamics the N$^3$LO  3NF component
is missing.  Therefore one
should take the determined values of the strengths with some caution.
 From the theoretical side, efforts to  include in the 3N continuum
  calculations consistently regularized
  N$^3$LO 3NF components are required what is the aim of
  the LENPIC collaboration. When such N$^3$LO 3NF becomes available
  the present study will be repeated.

  The elastic Nd scattering observables are driven by the S-matrix;
  therefore they are predominantly sensitive to the potential energy
  of three nucleons, whose main part is given by the pairwise interactions.
  Contrary to that, the nuclear
  bound   states  are sensitive to the interplay between the kinetic and potential
  energies of nucleons, being thus more sensitive to 3NF's. 
Based on the presented results it seems very probable that N$^4$LO contact
terms will have also large influence on spectra of nuclei. Therefore it
would be interesting to apply the nuclear Hamiltonian proposed in the
present paper to bound nuclear systems and see  what effects
the N$^4$LO contact terms have on the energy spectra and other properties of nuclei.

\begin{acknowledgments}
This study has been performed within Low Energy Nuclear Physics
International Collaboration (LENPIC) project.
 The numerical calculations were performed on the 
 supercomputer cluster of the JSC, J\"ulich, Germany.
\end{acknowledgments}

\appendix
\section{Momentum space partial wave decomposition of the N$^4$LO 3NF
  contact terms}
\label{app_a}

Here we introduce our definitions of the Faddeev components
corresponding to all the different N$^4$LO 3NF contact terms:
$E_2, E_3, E_4, E_5, E_6, E_8,
E_9, E_{10},$ and $E_{11}$.
Then we provide their 
momentum-space partial wave decomposition in the basis $\vert p~q~\alpha >$.
In the following 
$\vec p$ and $\vec q$ ($\vec {p}~'$ and $\vec {q}~'$)  
denote the relative initial (final) Jacobi momenta.
The other vectors,
$\vec q_i=\vec p_i~' - \vec p_i$ 
and $\vec K_i=\frac {\vec p_i~' + \vec p_i} {2}$,
are defined by the individual initial $\vec p_i$ (final $\vec p_i~'$) nucleon momenta ($i= 1, 2, 3$).

The partial wave decomposition for the $E_1$
and $E_7$ terms can be found in \cite{epel_tower}.
For details on our notation
we refer the reader to Ref.~\cite{glo96}. 
In particular we use $\hat X \equiv 2X+1$, where $X$ is an integer or a half-integer. 

For the E$_2$-term
\begin{eqnarray}
  V_{3N} &=& E_2 \sum_{i \ne j \ne k} \vec {q_i}^2 \vec {\tau}_i \cdot  \vec {\tau}_j
\label{eq_ap_4}
\end{eqnarray}
we define the Faddeev component as
\begin{eqnarray}
V_{3N}^{(1)} &=& E_2 \vec {q_1}^2 (\vec {\tau}_1 \cdot  \vec {\tau}_2  +
  \vec {\tau}_1 \cdot  \vec {\tau}_3 )
\label{eq_ap_5}
\end{eqnarray}  
and arrive at the following matrix elements:
\begin{eqnarray}
  <p'~q'~\alpha' \vert V_{3N}^{(1)} \vert p~q~\alpha > &=& \frac {1} {4\pi^4} E_2 
  \delta_{s's} \delta_{l'0} \delta_{l0}  \delta_{sj'} \delta_{sj} \delta_{T'T}
  \delta_{M_{T'}M_T} \delta_{t't} \cr
        && \times [ (q^2 + q'^2) \delta_{\lambda'0}  \delta_{\lambda0}
    \delta_{I' \frac {1} {2}} \delta_{I \frac {1} {2}} - \frac {2} {3} q q'
    \delta_{\lambda' 1} \delta_{\lambda 1} \delta_{I' I} ]  \cr
  && \times [-12 \sqrt{\hat t \hat t'} (-1)^{T-\frac {1} {2} }
\begin{Bmatrix}
      t' & t & 1 \\
      \frac {1} {2}  & \frac {1} {2} &  T
\end{Bmatrix}      
\begin{Bmatrix}
  t' & t & 1 \\
\frac {1} {2}   & \frac {1} {2} & \frac{1} {2}  
\end{Bmatrix}
                   ]  ~ .
\label{eq_ap_6}
\end{eqnarray}

For the E$_3$-term
\begin{eqnarray}
  V_{3N} &=& E_3 \sum_{i \ne j \ne k} \vec {q_i}^2 \vec {\sigma}_1` \cdot  \vec {\sigma}_j
\label{eq_ap_7}
\end{eqnarray}
we choose the Faddeev component as
\begin{eqnarray}
V_{3N}^{(1)} &=& E_3 \vec {q_1}^2 (\vec {\sigma}_i \cdot  \vec {\sigma}_2  +
  \vec {\sigma}_1 \cdot  \vec {\sigma}_3 )
\label{eq_ap_8}
\end{eqnarray}  
and obtain:
\begin{eqnarray}
  <p'~q'~\alpha' \vert V_{3N}^{(1)} \vert p~q~\alpha > &=&
  - \frac {1} {2\pi^4} 6 E_3 
  \delta_{s's} \delta_{l'0} \delta_{l0}  \delta_{sj'} \delta_{sj} \delta_{T'T}
  \delta_{M_{T'}M_T} \delta_{t't} \cr
  \times (-1)^{J-\frac {1} {2} } \hat s
\begin{Bmatrix}
  s & s & 1 \\
 \frac {1} {2}  & \frac {1} {2} & \frac {1} {2}  
\end{Bmatrix}
        &[& (q^2 + q'^2) \delta_{\lambda'0}  \delta_{\lambda0}
            \delta_{I' \frac {1} {2}} \delta_{I \frac {1} {2}}
\begin{Bmatrix}
  s & s & 1 \\
 \frac {1} {2}  & \frac {1} {2} & J   
\end{Bmatrix}
            \cr
             - \frac {2} {3} q q'
             \delta_{\lambda' 1}  \delta_{\lambda 1}  
             \sqrt{ \hat I \hat I' } &&
\begin{Bmatrix}
  1 & I' & I \\
 J  & j  &  j'
\end{Bmatrix}
\begin{Bmatrix}
 1 & I' & I \\
 1   & \frac {1} {2} &  \frac {1} {2}
\end{Bmatrix}
        ]  ~ .
\label{eq_ap_9}
\end{eqnarray}

For the E$_4$-term
\begin{eqnarray}
  V_{3N} &=& E_4 \sum_{i \ne j \ne k} \vec {q_i}^2 \vec {\sigma}_i \cdot
  \vec {\sigma}_j  \vec {\tau}_i \cdot  \vec {\tau}_j
\label{eq_ap_10}
\end{eqnarray}
we choose the Faddeev component in the following form:
\begin{eqnarray}
  V_{3N}^{(1)} &=& E_4 \vec {q_1}^2 (\vec {\sigma}_1 \cdot  \vec {\sigma}_2
   \vec {\tau}_1 \cdot  \vec {\tau}_2 +
  \vec {\sigma}_1 \cdot  \vec {\sigma}_3  \vec {\tau}_1 \cdot  \vec {\tau}_3  )
\label{eq_ap_11}
\end{eqnarray}  
and obtain
\begin{eqnarray}
  <p'~q'~\alpha' \vert V_{3N}^{(1)} \vert p~q~\alpha > &=&
  - \frac {1} {2\pi^4}  E_4 
  36 \sqrt{\hat s \hat s'}
\begin{Bmatrix}
s' & s & 1 \\
 \frac {1} {2}  & \frac {1} {2} &  \frac {1} {2}
\end{Bmatrix}
        \times (-1)^{J+\frac {1} {2} } \cr
         (-1)^{T' - \frac {1} {2} } && \sqrt{\hat t \hat t'}
\begin{Bmatrix}
 t' & t & 1 \\
 \frac {1} {2}  & \frac {1} {2} &  T
\end{Bmatrix}
\begin{Bmatrix}
 t' & t & 1 \\
 \frac {1} {2}  & \frac {1} {2} &  \frac {1} {2} 
\end{Bmatrix}
  \delta_{ll'} \delta_{l'0} \delta_{l0}  \delta_{sj'} \delta_{sj} \delta_{T'T}
  \delta_{M_{T'}M_T}  \cr
        &&\times [ (q^2 + q'^2) \delta_{\lambda'0}  \delta_{\lambda0}
    \delta_{I' \frac {1} {2}} \delta_{I \frac {1} {2}}
\begin{Bmatrix}
 s' & s & 1 \\
\frac {1} {2}   & \frac {1} {2} &  J
\end{Bmatrix}    
      \cr
          &&   - \frac {2} {3} q q'
      \delta_{\lambda' 1} \delta_{\lambda 1} \sqrt{ \hat I' \hat I }
      (-1)^{s'+s}
\begin{Bmatrix}
 I' & 1 & I  \\
  \frac {1} {2}  & 1 &  \frac {1} {2}
\end{Bmatrix}
\begin{Bmatrix}
 I' & 1 & I  \\  
 j   & J &  j'
\end{Bmatrix}  
 ]  ~ .
\label{eq_ap_12}
\end{eqnarray}

For the E$_5$-term
\begin{eqnarray}
  V_{3N} &=& E_5 \sum_{i \ne j \ne k} ( 3 \vec q_i \cdot \vec {\sigma}_i
  \vec q_i \cdot \vec {\sigma}_j
  - \vec {q_i}^2 \vec {\sigma}_i \cdot
  \vec {\sigma}_j  )
\label{eq_ap_13}
\end{eqnarray}
our definition of the Faddeev component is
\begin{eqnarray}
  V_{3N}^{(1)} &=& E_5 [ 3 \vec {q_1} \cdot  \vec {\sigma}_1
    ( \vec {q_1} \cdot  \vec {\sigma}_2 + \vec {q_1} \cdot  \vec {\sigma}_3 ) 
   - \vec {q_1}^2 (\vec {\sigma}_1 \cdot  \vec {\sigma}_2 + 
   \vec {\sigma}_1 \cdot  \vec {\sigma}_3 ) ]
\label{eq_ap_14}
\end{eqnarray}  
and we get:
\begin{eqnarray}
  <p'~q'~\alpha' \vert V_{3N}^{(1)} \vert p~q~\alpha > &=&
   \frac {1} {2\pi^4}  E_5 
   6 \sqrt{\hat s \hat s'}
\begin{Bmatrix}
 s' & s & 1 \\
 \frac {1} {2}  & \frac {1} {2} &  \frac {1} {2}
\end{Bmatrix}
      \delta_{l'0} \delta_{l0}  \delta_{j's'} \delta_{js} \delta_{s's}
      \delta_{tt'} \delta_{T'T}
  \delta_{M_{T'}M_T}  \cr
  \times
      [ 3q'q'\sqrt{ \frac {\hat I'}  {\hat J} } (-1)^{j' + I' +s} \delta_{\lambda 0}
      \delta_{I \frac {1} {2} } && <1010\vert \lambda' 0>
      \sum_{S'}{\hat S'\sqrt{\hat S'}(-1)^{S'}
\begin{Bmatrix}
 J & I' & j'  \\
 \frac {1} {2}   & S' &  \lambda'
\end{Bmatrix}        
\begin{Bmatrix}
 S' & J & \lambda'  \\
 s' & s & 1 \\
\frac {1} {2}   & \frac {1} {2} &  1 
\end{Bmatrix}
 }    
      \cr
     +3qq\sqrt{  {\hat I}  {\hat J} } (-1)^{j + I +s'} \delta_{\lambda' 0}
      \delta_{I' \frac {1} {2} } && <1010\vert \lambda 0>
      \sum_{S} \sqrt{ {\hat S} } (-1)^{S}
\begin{Bmatrix}
 J & I & j \\
\frac {1} {2}   & S &  \lambda 
\end{Bmatrix}
\begin{Bmatrix}
 S & J & \lambda \\
 s & s' & 1 \\
 \frac {1} {2}   & \frac {1} {2} &  1 
\end{Bmatrix}
      \cr
      -2q'q\sqrt{  {\hat I}  {\hat I'} } (-1)^{J + \frac {1} {2} +j'}
      \delta_{\lambda' 1}
      \delta_{\lambda 1 } &&
\begin{Bmatrix}
 I' & 1 & I  \\
 \frac {1} {2}   & 1 &  \frac {1} {2}         
\end{Bmatrix}
\begin{Bmatrix}
  I' & 1 & I   \\
  j   & J &  j' 
\end{Bmatrix}  
            \cr
    &&  + (-1)^{J+ \frac {1} {2} } \{  (q^2 + q'^2)
        \delta_{\lambda'0}  \delta_{\lambda0}
        \delta_{I' \frac {1} {2}} \delta_{I \frac {1} {2}}
\begin{Bmatrix}
 s' & s & 1 \\
 \frac {1} {2}   & \frac {1} {2} &  J          
\end{Bmatrix}
      \cr
          &&   - \frac {2} {3} q q'
      \delta_{\lambda' 1} \delta_{\lambda 1} \sqrt{ \hat I' \hat I }
\begin{Bmatrix}
 I' & 1 & I  \\
 \frac {1} {2}  & 1 &  \frac {1} {2}
\end{Bmatrix}      
\begin{Bmatrix}
  I' & 1 & I  \\
j   & J &  j'  
\end{Bmatrix}
 \} ]  ~ .
\label{eq_ap_15}
\end{eqnarray}

For the E$_6$-term
\begin{eqnarray}
  V_{3N} &=& E_6 \sum_{i \ne j \ne k} ( 3 \vec q_i \cdot \vec {\sigma}_i
  \vec q_i \cdot \vec {\sigma}_j
  - \vec {q_i}^2 \vec {\sigma}_i \cdot
  \vec {\sigma}_j  )  \vec {\tau}_i \cdot  \vec {\tau}_j
\label{eq_ap_16}
\end{eqnarray}
our choice of the Faddeev component reads
\begin{eqnarray}
  V_{3N}^{(1)} &=& E_6 [ ( 3 \vec {q_1} \cdot  \vec {\sigma}_1
                           \vec {q_1} \cdot  \vec {\sigma}_2   
   - \vec {q_1}^2 \vec {\sigma}_1 \cdot  \vec {\sigma}_2 )  
   \vec {\tau}_1 \cdot  \vec {\tau}_2 \cr
 && +( 3 \vec {q_1} \cdot  \vec {\sigma}_1
                           \vec {q_1} \cdot  \vec {\sigma}_3   
   - \vec {q_1}^2 \vec {\sigma}_1 \cdot  \vec {\sigma}_3 )  
   \vec {\tau}_1 \cdot  \vec {\tau}_3
  ]
\label{eq_ap_17}
\end{eqnarray}  
and we get:
\begin{eqnarray}
  <p'~q'~\alpha' \vert V_{3N}^{(1)} \vert p~q~\alpha > &=&
  - \frac {1} {2\pi^4} E_6 
  36 \sqrt{\hat j \hat j'}
\begin{Bmatrix}
 j' & j & 1  \\
  \frac {1} {2}  & \frac {1} {2} &  \frac {1} {2}   
\end{Bmatrix}
    \delta_{ll'}   \delta_{l'0} \delta_{l0}  \delta_{s'j'} \delta_{sj}  \cr
  \times
      [ 3q'q'\sqrt{ \frac {\hat I'}  {\hat J} } (-1)^{j'+j + I'} \delta_{\lambda 0}
      \delta_{I \frac {1} {2} } && <1010\vert \lambda' 0>
      \sum_{S'}{\hat S'\sqrt{\hat S'}(-1)^{S'}
 \begin{Bmatrix}
 J & I' & j'  \\
 \frac {1} {2}   & S' &  \lambda'
 \end{Bmatrix}       
 \begin{Bmatrix}
   S' & J & \lambda' \\
 \frac {1} {2}   & \frac {1} {2} &  1    
 \end{Bmatrix}
  }    
      \cr
     +3qq\sqrt{  {\hat I}  {\hat J} } (-1)^{j'+j + I} \delta_{\lambda' 0}
      \delta_{I' \frac {1} {2} } && <1010\vert \lambda 0>
      \sum_{S}{ \sqrt{ \hat S } (-1)^{S}
 \begin{Bmatrix}
 J & I & j  \\
 \frac {1} {2}   & S &  \lambda
 \end{Bmatrix}        
 \begin{Bmatrix}
   S & J & \lambda \\
   s & s' & 1 \\
 \frac {1} {2}   & \frac {1} {2} &  1   
 \end{Bmatrix}
  }    
      \cr
      -2q'q\sqrt{  {\hat I}  {\hat I'} } (-1)^{J + \frac {1} {2}+ j'}
      \delta_{\lambda' 1}
      \delta_{\lambda 1 } &&
\begin{Bmatrix}
 I' & 1 & I  \\
 \frac {1} {2}   & 1 &  \frac {1} {2} 
\end{Bmatrix}
\begin{Bmatrix}
  I' & 1 & I \\
  j   & J &  j' 
\end{Bmatrix}  
      \cr
      -  (-1)^{J+ \frac {1} {2} } && \{  (q^2 + q'^2)
        \delta_{\lambda'0}  \delta_{\lambda0}
        \delta_{I' \frac {1} {2}} \delta_{I \frac {1} {2}}
\begin{Bmatrix}
 s' & s & 1 \\
 \frac {1} {2}   & \frac {1} {2} &  J
\end{Bmatrix}
              \cr
             - \frac {2} {3} q q'
      \delta_{\lambda' 1} \delta_{\lambda 1} \sqrt{ \hat I' \hat I }
      && (-1)^{j+j'}
 \begin{Bmatrix}
 I' & 1 & I  \\
 \frac {1} {2}  & 1 &  \frac {1} {2}
\end{Bmatrix}
 \begin{Bmatrix}
 I' & 1 & I  \\
  j   & J &  j'     
 \end{Bmatrix}
 \} ] \cr
      \times [  \delta_{T'T}  \delta_{M_{T'}M_T} \sqrt{\hat t \hat t'}
        &&   (-1)^{T'-\frac {1} {2} }
 \begin{Bmatrix}
 t' & t & 1  \\
 \frac {1} {2}   & \frac {1} {2} & T
 \end{Bmatrix}        
 \begin{Bmatrix}
 t' & t & 1   \\
  \frac {1} {2}   & \frac {1} {2} & 1  
 \end{Bmatrix} 
      ]  ~ .
\label{eq_ap_18}
\end{eqnarray}

For the E$_8$-term
\begin{eqnarray}
  V_{3N} &=& i E_8 \sum_{i \ne j \ne k} \vec {q_i} \times (\vec {K}_i - \vec {K}_j)
  \cdot (\vec {\sigma}_i + \vec {\sigma}_j )
  \vec {\tau}_j  \cdot \vec {\tau}_k
  \label{eq_ap_22}
\end{eqnarray}
we choose the Faddeev component as
\begin{eqnarray}
V_{3N}^{(1)} &=& i E_8 [ \vec {q_1} \times (\vec {K}_1 - \vec {K}_2)
  \cdot (\vec {\sigma}_1 + \vec {\sigma}_2)
  \vec {\tau}_2  \cdot \vec {\tau}_3  +
\vec {q_1} \times (\vec {K}_1 - \vec {K}_3)
\cdot (\vec {\sigma}_1 + \vec {\sigma}_3 )
\vec {\tau}_3  \cdot \vec {\tau}_2 ]
\label{eq_ap_23}
\end{eqnarray}
and get:
\begin{eqnarray}
  <p'~q'~\alpha' \vert V_{3N}^{(1)} \vert p~q~\alpha > &=&
  - \frac {1} {8\pi^4}  E_8
  \delta_{T'T} \delta_{M_{T'}M_T} \delta_{t't} 6(-1)^t
\begin{Bmatrix}
  \frac {1} {2} &  \frac {1} {2} & 1  \\
   \frac {1} {2}  & \frac {1} {2} &  t
\end{Bmatrix}
    \cr
  &&\times [ \, - \sqrt {2} (1-(-1)^{s'+s}) (-1)^{J+\frac {1} {2}} \cr
      &&\times ( \, ~  qp \delta_{l' 0}  \delta_{\lambda' 0} \delta_{l 1}
      \delta_{\lambda 1} \delta_{s' j'} \delta_{I' \frac {1} {2}}
      \sqrt{\hat j \hat I}
\begin{Bmatrix}
  1 & s & s'  \\
   j  & 1 &  1
\end{Bmatrix}      
\begin{Bmatrix}
  s' & j & 1 \\
 I  & \frac {1} {2} &  J  
\end{Bmatrix}  
 \cr
      && ~ -qp' \delta_{l' 1}  \delta_{\lambda' 0} \delta_{l 0}
      \delta_{\lambda 1} \delta_{s j} \delta_{I' \frac {1} {2}}
      \sqrt{\hat j' \hat I}
\begin{Bmatrix}
 1 & s' & s  \\
 j'  & 1 &  1
\end{Bmatrix}
\begin{Bmatrix}      
 j' & s & 1  \\
 I  & \frac {1} {2} &  J
\end{Bmatrix}
 \cr
      && ~ -q'p \delta_{l' 0}  \delta_{\lambda' 1} \delta_{l 1}
      \delta_{\lambda 0} \delta_{s' j'} \delta_{I \frac {1} {2}}
      \sqrt{\hat j \hat I'}
\begin{Bmatrix} 
 1 & s & s'  \\
j  & 1 &  1
\end{Bmatrix}
\begin{Bmatrix}
 j & s' & 1  \\
 I'  & \frac {1} {2} &  J
\end{Bmatrix}
 \cr
      && ~ +q'p' \delta_{l' 1}  \delta_{\lambda' 1} \delta_{l 0}
      \delta_{\lambda 0} \delta_{s j} \delta_{I \frac {1} {2}}
      \sqrt{\hat j' \hat I'}
\begin{Bmatrix}
  1 & s' & s  \\
  j'  & 1 &  1  
\end{Bmatrix}        
\begin{Bmatrix}
  s & j' & 1   \\
 I'  & \frac {1} {2} &  J 
\end{Bmatrix}   
       ) \, \cr
      && ~ +12qq' \delta_{l' 0}  \delta_{\lambda' 1} \delta_{l 0} 
      \delta_{\lambda 1} \delta_{s' s} \delta_{s' j'} \delta_{s j}
      ( \delta_{I'I} (-1)^{I+ \frac {1} {2} }
\begin{Bmatrix}
\frac {1} {2}  & 1 & I  \\
 1  & \frac {1} {2} &  1
\end{Bmatrix}       
\cr
&& + \delta_{s1} \sqrt{\hat I \hat I'} (-1)^{I'+I+J+\frac {1} {2}}
\begin{Bmatrix}
 1  & I & I'  \\
 \frac {1} {2}  & 1 &  1
\end{Bmatrix}
\begin{Bmatrix}
  1  & I & I'  \\
   J  & 1 &  1
\end{Bmatrix}
   )        
  ] \, ~ .
\label{eq_ap_24}
\end{eqnarray}

For the E$_9$-term
\begin{eqnarray}
  V_{3N} &=& E_9 \sum_{i \ne j \ne k}   \vec q_i \cdot \vec {\sigma}_i
  \vec q_j \cdot \vec {\sigma}_j
\label{eq_ap_25}
\end{eqnarray}
we define the Faddeev component as
\begin{eqnarray}
  V_{3N}^{(1)} &=& E_9 [  \vec {q_1} \cdot  \vec {\sigma}_1
    ( \vec {q_2} \cdot  \vec {\sigma}_2 + \vec {q_3} \cdot  \vec {\sigma}_3 ) ] 
\label{eq_ap_26}
\end{eqnarray}  
and get:
\begin{eqnarray}
  <p'~q'~\alpha' \vert V_{3N}^{(1)} \vert p~q~\alpha > &=&
   \frac {1} {2\pi^4} E_9 
   \delta_{tt'} \delta_{T'T} \delta_{M_{T'}M_T}
\begin{Bmatrix}
  s' & s & 1  \\
\frac {1} {2}  & \frac {1} {2} &  \frac {1} {2}  
\end{Bmatrix}
       \sqrt{\hat s \hat s'} \cr
  \times
      [ ( \frac {1-(-1)^{s+s'}} {2} ) \{ &-& \sqrt{2} q'p
        \delta_{l'0} \delta_{\lambda' 1} \delta_{l1} \delta_{\lambda 0} \delta_{j's'}
        \delta_{I \frac {1} {2} }
        \sqrt{  {\hat j\hat I}  {\hat I'} } (-1)^{j' + I'+J +s}  \cr
        \sum_{S}{ \sqrt{\hat S}} &&
\begin{Bmatrix}
 s & j & 1 \\
 J   & S &  \frac {1} {2}
\end{Bmatrix}        
\begin{Bmatrix}
  \frac {1} {2} & s & S  \\
  J & I' & j' 
\end{Bmatrix}
\begin{Bmatrix}
  J & I' & j'  \\
  \frac {1} {2}   & S & 1
\end{Bmatrix}
        \cr
     &-& \sqrt{2} qp'
        \delta_{l'1} \delta_{\lambda' 0} \delta_{l0} \delta_{\lambda 1} \delta_{js}
        \delta_{I' \frac {1} {2} }
        \sqrt{  \hat j' \hat I  \hat I' } (-1)^{j + I + J + s'}  \cr
        \sum_{S}{ \sqrt{\hat S} } &&
\begin{Bmatrix}
 s' & j' & 1  \\
J   & S &  \frac {1} {2}
\end{Bmatrix}        
\begin{Bmatrix}
  \frac {1} {2} & s & S \\
   s' & \frac {1} {2} & 1
\end{Bmatrix}
\begin{Bmatrix}
  J & I & j  \\
  \frac {1} {2}   & S & 1 
\end{Bmatrix}
        \cr
     &+& 2 qp 
        \delta_{l'0} \delta_{\lambda' 0} \delta_{l1} \delta_{\lambda 1} \delta_{j's'}
        \delta_{I' \frac {1} {2} }
        \sqrt{  {\hat j \hat I}  {\hat J} } (-1)^{ I' - \frac {1} {2} - J - s}  \cr
        &&    \sum_{L,S}{ \hat L \sqrt{\hat S}  (-1)^{L+S} }
\begin{Bmatrix}
 1 & s & j  \\
 1   & \frac {1} {2} & I  \\
  L   & S & J
\end{Bmatrix}        
\begin{Bmatrix}
  J & S & L \\
   s' & s & 1  \\
 \frac {1} {2}  &  \frac {1} {2} & 1
\end{Bmatrix}
        \cr
     &+& 2 q'p' 
        \delta_{l'1} \delta_{\lambda' 1} \delta_{l0} \delta_{\lambda 0} \delta_{js}
        \delta_{I \frac {1} {2} }
        \sqrt{  {\hat j' \hat I'}  } (-1)^{ I - \frac {1} {2} - J + s'}  \cr
        \sum_{L',S'}{ \hat L' \hat S'} &&  (-1)^{L'+S'}
\begin{Bmatrix}
  1 & s' & j'  \\
  1 & \frac {1} {2} & I'  \\
 L'   & S' & J
\end{Bmatrix}
\begin{Bmatrix}        
  J & S' & L'  \\
  s & s' & 1  \\
  \frac {1} {2} & \frac {1} {2} & 1  
\end{Bmatrix}
\}
        \cr
 + 
  ( \frac {1+(-1)^{s+s'}} {2} ) \{ -3q'q'  \delta_{l 0} \delta_{l' 0} &&
   \delta_{\lambda 0} \delta_{js} \delta_{j's' }  \delta_{I \frac {1} {2} }
    \sqrt{ \frac  {\hat I'}  {\hat J} } (-1)^{j' - I' +  s +1} 
 <1010\vert \lambda' 0> \cr
 &&    \sum_{S'}{\hat S'\sqrt{\hat S'}(-1)^{S'}
\begin{Bmatrix}
  J & I' & j'  \\
  \frac {1} {2}   & S' &  \lambda' 
\end{Bmatrix} 
\begin{Bmatrix}
 S' & J & \lambda'  \\   
 s' & s & 1  \\
 \frac {1} {2}   & \frac {1} {2} &  1 
\end{Bmatrix}
  }    
      \cr
     - 3qq \delta_{l 0} \delta_{l' 0} &&
     \delta_{\lambda' 0} \delta_{js} \delta_{j's' }  \delta_{I' \frac {1} {2} }
     \sqrt{  {\hat I}  {\hat J} } (-1)^{j - I + s' + 1} 
      <1010\vert \lambda 0>   \cr
      &&   \sum_{S}{\hat S (-1)^{S}
\begin{Bmatrix}
 J & I & j  \\
 \frac {1} {2}   & S &  \lambda
\end{Bmatrix}
\begin{Bmatrix}        
S & J & \lambda   \\
s & s' & 1  \\
   \frac {1} {2}   & \frac {1} {2} &  1
\end{Bmatrix}
  }    
      \cr
      +2q'q  \delta_{l 0} \delta_{l' 0}
     \delta_{\lambda 1} \delta_{\lambda' 1}   \delta_{js} \delta_{j's' } 
      \sqrt{  {\hat I}  {\hat I'} } (-1)^{J + \frac {1} {2} +j' } 
      &&
\begin{Bmatrix}
  I' & 1 & I \\
\frac {1} {2}   & 1 &  \frac {1} {2}  
\end{Bmatrix}
\begin{Bmatrix}
 I' & 1 & I  \\
 j   & J &  j' 
\end{Bmatrix}
      \} ]  ~ .
\label{eq_ap_27}
\end{eqnarray}

For the E$_{10}$-term
\begin{eqnarray}
  V_{3N} &=& E_{10} \sum_{i \ne j \ne k}   \vec q_i \cdot \vec {\sigma}_i
  \vec q_j \cdot \vec {\sigma}_j  \vec {\tau}_i \cdot \vec {\tau}_j
\label{eq_ap_28}
\end{eqnarray}
we choose the Faddeev component
\begin{eqnarray}
  V_{3N}^{(1)} &=& E_{10}   \vec {q_1} \cdot  \vec {\sigma}_1 [
     \vec {q_2} \cdot  \vec {\sigma}_2 \vec {\tau}_1 \cdot \vec {\tau}_2
    + \vec {q_3} \cdot  \vec {\sigma}_3   \vec {\tau}_1 \cdot \vec {\tau}_3 ]
\label{eq_ap_29}
\end{eqnarray}  
and get:
\begin{eqnarray}
  <p'~q'~\alpha' \vert V_{3N}^{(1)} \vert p~q~\alpha > &=&
  \frac {1} {2\pi^4} E_{10}
\begin{Bmatrix}
 s' & s & 1  \\
\frac {1} {2}  & \frac {1} {2} &  \frac {1} {2}
\end{Bmatrix}  
      \sqrt{\hat s \hat s'} \cr
      \times 
      (-6  \delta_{T'T} \delta_{M_{T'}M_{T}}
      && \sqrt{\hat t \hat t'} (-1)^{T-\frac {1} {2} }
\begin{Bmatrix}
  t' & t & 1  \\
  \frac {1} {2}  & \frac {1} {2} &  T  
\end{Bmatrix}
\begin{Bmatrix}
 t' & t & 1  \\
 \frac {1} {2}  & \frac {1} {2} &   \frac {1} {2} 
\end{Bmatrix}
   )  
      \cr
  \times
      [ ( \frac {1-(-1)^{l+l'}} {2} ) \{ &-& \sqrt{2} q'p
        \delta_{l'0} \delta_{\lambda' 1} \delta_{l1} \delta_{\lambda 0} \delta_{j's'}
        \delta_{I \frac {1} {2} }
        \sqrt{  {\hat j\hat I}  {\hat I'} } (-1)^{j' + I'+J +s}  \cr
        &&   \sum_{S}{ \sqrt{\hat S}
\begin{Bmatrix}
 s & j & 1  \\
 J   & S &  \frac {1} {2}
\end{Bmatrix} 
\begin{Bmatrix}          
  \frac {1} {2} & s & S \\
  s' & \frac {1} {2} & 1  
\end{Bmatrix}
\begin{Bmatrix}
 J & I' & j'  \\
 \frac {1} {2}   & S & 1
\end{Bmatrix} 
  }   
        \cr
     &-& \sqrt{2} qp'
        \delta_{l'1} \delta_{\lambda' 0} \delta_{l0} \delta_{\lambda 1} \delta_{js}
        \delta_{I' \frac {1} {2} }
        \sqrt{  \hat j' \hat I  \hat I' } (-1)^{j + I + J + s'}  \cr
        &&   \sum_{S}{ \sqrt{\hat S}
\begin{Bmatrix}
 s' & j' & 1  \\
 J   & S &  \frac {1} {2}
\end{Bmatrix}
\begin{Bmatrix}          
  \frac {1} {2} & s & S  \\
   s' & \frac {1} {2} & 1
\end{Bmatrix}
\begin{Bmatrix} 
 J & I & j  \\
 \frac {1} {2}   & S & 1
\end{Bmatrix}
 }   
        \cr
     &+& 2 qp 
        \delta_{l'0} \delta_{\lambda' 0} \delta_{l1} \delta_{\lambda 1} \delta_{j's'}
        \delta_{I' \frac {1} {2} }
        \sqrt{  {\hat j \hat I}  {\hat J} } (-1)^{ I' - \frac {1} {2} - J - s}  \cr
        &&    \sum_{L,S}{ \hat L \sqrt{\hat S}  (-1)^{L+S} }
\begin{Bmatrix} 
 1 & s & j \\
 1   & \frac {1} {2} & I  \\
  L   & S & J
\end{Bmatrix}
\begin{Bmatrix}        
  J & S & L  \\
  s' & s & 1  \\
  \frac {1} {2}  &  \frac {1} {2} & 1 
\end{Bmatrix}
     \cr
     &+& 2 q'p' 
        \delta_{l'1} \delta_{\lambda' 1} \delta_{l0} \delta_{\lambda 0} \delta_{js}
        \delta_{I \frac {1} {2} }
        \sqrt{  {\hat j' \hat I'}  } (-1)^{ I - \frac {1} {2} - J + s'}  \cr
        \sum_{L',S'}{ \hat S' \hat L'} && (-1)^{L'+S'}
 \begin{Bmatrix}
          1 & s' & j'  \\
 1 & \frac {1} {2} & I'  \\
 L'   & S' & J          
\end{Bmatrix}
 \begin{Bmatrix}       
   J & S' & L'  \\
    s & s' & 1  \\
\frac {1} {2} & \frac {1} {2} & 1
\end{Bmatrix}
          \}
        \cr
 + 
  ( \frac {1+(-1)^{l+l'}} {2} ) \{ -3q'q'  \delta_{l 0} \delta_{l' 0} &&
   \delta_{\lambda 0} \delta_{js} \delta_{j's' }  \delta_{I \frac {1} {2} }
    \sqrt{ \frac  {\hat I'}  {\hat J} } (-1)^{j' - I' +  s +1} 
 <1010\vert \lambda' 0> \cr
 &&    \sum_{S'}{\hat S'\sqrt{\hat S'}(-1)^{S'}
\begin{Bmatrix}
  J & I' & j' \\
 \frac {1} {2}   & S' &  \lambda' 
\end{Bmatrix}
\begin{Bmatrix}
  S' & J & \lambda'  \\
  s' & s & 1 \\
\frac {1} {2}   & \frac {1} {2} &  1  
\end{Bmatrix}
  }    
      \cr
     - 3qq \delta_{l 0} \delta_{l' 0} &&
     \delta_{\lambda' 0} \delta_{js} \delta_{j's' }  \delta_{I' \frac {1} {2} }
     \sqrt{  {\hat I}  {\hat J} } (-1)^{j - I + s' + 1} 
      <1010\vert \lambda 0>   \cr
      &&   \sum_{S}{\hat S (-1)^{S}
\begin{Bmatrix}
  J & I & j   \\
\frac {1} {2}   & S &  \lambda  
\end{Bmatrix}
\begin{Bmatrix}
  S & J & \lambda  \\       
  s & s' & 1  \\
 \frac {1} {2}   & \frac {1} {2} &  1  
\end{Bmatrix}
  }    
      \cr
      +2q'q  \delta_{l 0} \delta_{l' 0}
     \delta_{\lambda 1} \delta_{\lambda' 1}   \delta_{js} \delta_{j's' } 
      \sqrt{  {\hat I}  {\hat I'} } (-1)^{J + \frac {1} {2} +j'} 
      &&
\begin{Bmatrix}
  I' & 1 & I  \\
 \frac {1} {2}   & 1 &  \frac {1} {2} 
\end{Bmatrix}      
\begin{Bmatrix}
  I' & 1 & I  \\
  j   & J &  j' 
\end{Bmatrix}
      \} ]  ~ .
\label{eq_ap_30}
\end{eqnarray}

There are three additional contact terms coming with strengths
 $E_{11}$, $E_{12}$, and
$E_{13}$. 
For the E$_{11}$-term
\begin{eqnarray}
  V_{3N} &=& E_{11} \sum_{i \ne j \ne k}   \vec q_i \cdot \vec {\sigma}_j
  \vec q_j \cdot \vec {\sigma}_i  
\label{eq_ap_31}
\end{eqnarray}
we choose the Faddeev component as
\begin{eqnarray}
  V_{3N}^{(1)} &=& E_{11} [  \vec {q_1} \cdot  \vec {\sigma}_2 
     \vec {q_2} \cdot  \vec {\sigma}_1 
    + \vec {q_1} \cdot  \vec {\sigma}_3   \vec {q_3} \cdot  \vec {\sigma}_1 ]
\label{eq_ap_32}
\end{eqnarray}

For the E$_{12}$-term
\begin{eqnarray}
  V_{3N} &=& E_{12} \sum_{i \ne j \ne k}   \vec q_i \cdot \vec {\sigma}_j
  \vec q_j \cdot \vec {\sigma}_i   \vec {\tau}_i \cdot \vec {\tau}_j 
\label{eq_ap_33}
\end{eqnarray}
we choose the Faddeev component
\begin{eqnarray}
  V_{3N}^{(1)} &=& E_{12} [  \vec {q_1} \cdot  \vec {\sigma}_2 
     \vec {q_2} \cdot  \vec {\sigma}_1 \vec {\tau}_1 \cdot \vec {\tau}_2 
     + \vec {q_1} \cdot  \vec {\sigma}_3   \vec {q_3} \cdot  \vec {\sigma}_1 
  \vec {\tau}_1 \cdot \vec {\tau}_3 ]
\label{eq_ap_34}
\end{eqnarray}  
    
For the E$_{13}$-term
\begin{eqnarray}
  V_{3N} &=& E_{13} \sum_{i \ne j \ne k}   \vec q_i \cdot \vec {\sigma}_j
  \vec q_j \cdot \vec {\sigma}_i   \vec {\tau}_i \cdot \vec {\tau}_k 
\label{eq_ap_35}
\end{eqnarray}
we choose the Faddeev component
\begin{eqnarray}
  V_{3N}^{(1)} &=& E_{13} [  \vec {q_1} \cdot  \vec {\sigma}_2 
     \vec {q_2} \cdot  \vec {\sigma}_1 \vec {\tau}_1 \cdot \vec {\tau}_3 
     + \vec {q_1} \cdot  \vec {\sigma}_3   \vec {q_3} \cdot  \vec {\sigma}_1 
  \vec {\tau}_1 \cdot \vec {\tau}_2 ]
\label{eq_ap_36}
\end{eqnarray}  

The partial wave decomposition of the $E_{11}$ term is identical with 
the $E_{9}$ term and of the $E_{12}$ term with the
 $E_{10}$ term. The partial wave
decomposition of the $E_{13}$ term differs from that of the $E_{10}$ term only
 by a factor of  $(-1)^{t+t'}$.

\newpage

\begin{table}[ht] 
  \caption{Contributions of the N$^2$LO and N$^4$LO contact terms to
    the potential energy of the 
    three nucleons in the triton. These expectation values were obtained
    for the  $^3$H wave function calculated with the SMS chiral N$^4$LO$^+$
    NN potential ($\Lambda=450$~MeV)  and
    assuming strengths of contact terms $c_{i}=1.0$.
\label{tab1}}
\begin{center}
\begin{tabular}{lcc}
\hline
 $V_i$   &  $< \psi_{^3H} \vert V_{i} \vert \psi_{^3H}>$      \\
&       [MeV]      \\
\hline
$V_D$     &  0.1661   \\
$V_E$     &  -1.4294  \\
$V_{E1}$    &   0.3463  \\
$V_{E2}$ &      -0.4173 \\
$V_{E3}$ &  -0.2754     \\
$V_{E4}$ & -1.0390    \\
$V_{E5}$ &  -0.9559  \\
$V_{E6}$ & -1.0699   \\
$V_{E7}$ &  0.1798 $\times 10^{-4}$  \\
$V_{E8}$ &  0.8817 $\times 10^{-2}$    \\
$V_{E9}$ & -0.2407   \\
$V_{E10}$  & 1.0571    \\
$V_{E11}$  & -0.2407   \\
$V_{E12}$  &   1.0571        \\
$V_{E13}$  &   0.3060     \\[4pt]
\hline
\end{tabular}
\end{center}
\end{table}
\newpage

\begin{table}
  \caption{The data basis used  for fixing the strengths of the contact terms $c_i$.}
\label{tab2} 
\begin{tabular}{|c|c|c|c|c|c|c|}
\hline
$E$ & $\frac {d\sigma} {d\Omega}$ & $A_y$ & $iT_{11}$ & $T_{20}$ &  $T_{21}$ & $T_{22}$  \cr
[MeV]  &    &    &  &  &   &   \cr
\hline
10 & nd~\cite{how_sig10}, pd~\cite{sagara} &  nd~\cite{how_ay10},
pd~\cite{sagara,sperisen84} & pd~\cite{sperisen84,sawada83} & pd~\cite{sperisen84}  &
pd~\cite{sperisen84}  & pd~\cite{sperisen84}   \\
\hline
70 & pd~\cite{seki140} & pd~\cite{shimizu} ($65$~MeV)  & pd~\cite{seki140} & pd~\cite{seki140} & pd~\cite{seki140} & pd~\cite{seki140}  \\
\hline
135 & pd~\cite{seki140,sak99}  & pd~\cite{ermisch,iucf2006} & pd~\cite{seki140} &
pd~\cite{seki140} &  pd~\cite{seki140} & pd~\cite{seki140}  \\
\hline
190 & pd~\cite{ermisch}  & pd~\cite{ermisch} & pd~\cite{seki190} &
pd~\cite{seki190} & pd~\cite{seki190}  & pd~\cite{seki190}  \\
\hline
250 & nd~\cite{maeda250}, pd~\cite{hatanaka}  & pd~\cite{hatanaka} &
pd~\cite{seki250} & pd~\cite{seki250} & pd~\cite{seki250} & pd~\cite{seki250}  \\
\hline
\end{tabular}
\end{table}

\newpage

\begin{table}[ht] 
  \caption{The values of strengths  $c_i$ found in the least-squares fit 
    to the data from Table.~\ref{tab2} at three
    energies $E=10, 70$, and $135$~MeV.
\label{tab3}}
\begin{center}
\begin{tabular}{lcc}
%
%
%
\hline
$c_D$  & &  -1.49 $\pm$  0.06  \\
$c_E$  & &  -1.27 $\pm$  0.06  \\
$c_{E_1}$ & &   6.40 $\pm$  0.33  \\
$c_{E_2}$ & &   7.80 $\pm$  0.36   \\
$c_{E_3}$ & &   6.97 $\pm$  0.34  \\ 
$c_{E_4}$ & &  -2.06 $\pm$  0.13 \\
$c_{E_5}$ & &  -0.36 $\pm$  0.05  \\
$c_{E_6}$ & &   0.52  $\pm$  0.03    \\
$c_{E_7}$ & &  -7.40 $\pm$  0.14   \\
$c_{E_8}$ & &  -2.61 $\pm$  0.05  \\
$c_{E_9}$ & &  -4.59 $\pm$  0.22  \\
$c_{E_{10}}$ & &  -0.98 $\pm$  0.05  \\
$c_{E_{13}}$ & &  -1.14 $\pm$  0.05  \\ [4pt]
\hline
\end{tabular}
\end{center}
\end{table}
\newpage

\begin{table}[ht] 
  \caption{The covariance matrix for the strengths $c_i$ determined by the
    the least-squares fit of data from Table~\ref{tab2} 
    at three energies $E=10, 70,$ and $135$~MeV (the
    values shown are Cov($c_{i},c_j$) $\times 1000$).
\label{tab4}}
\begin{center}
\begin{tabular}{lccccccccccccc}
\hline
 & $c_D$ & $c_E$ & $c_{E_1}$ & $c_{E_2}$ & $c_{E_3}$ & $c_{E_4}$
& $c_{E_5} $ & $c_{E_6}$ & $c_{E_7}$ & $c_{E_8}$ & $c_{E_9}$ & $c_{E_{10}}$
& $c_{E_{13}}$ \\
\hline
$c_D$  &    3.914   & -0.456 & 1.412 & 4.573 & 0.843 & 0.844 & -0.729 &  -0.892 & 1.109 & 0.267 & -0.726 & 0.123 & -0.207 \\
$c_E$  &    & 3.560 & 0.947 & -3.571 & 1.345  & -0.633 & -0.172 & -0.217 & -2.416 & -0.809  & -1.702 & 0.393  &  0.571 \\
$c_{E_1}$ &   & & 108.9 & 112.8 & 108.9 & -35.13 & 1.409 & -2.418 & 25.92 & 7.513 & 12.99 & 3.861  &  0.443 \\
$c_{E_2}$ &   & & &  130.7 &  113.4  & -35.15 & -1.995 & -3.241 & 32.43 & 9.561  & -0.534 & 0.763 & -3.332 \\
$c_{E_3}$ &  & & & & 112.9 & -38.92 & 1.617 & -1.814 & 27.52  & 8.068 &  8.366 &  1.598 & -0.193 \\
$c_{E_4}$ &  & & & & & 15.97 & -1.966 & -0.362 & -10.50 & -3.198 & -4.866 & 0.345 & -0.222 \\
$c_{E_5}$ &  & & & & & & 2.415 & 0.669 & 0.791 & 0.281 &  9.892 & 1.311  &  1.766 \\
$c_{E_6}$ &  & & & & & & & 0.635 & -0.874 & -0.226 & 1.426 & -0.226 & 0.210 \\
$c_{E_7}$ &  & & & & & & & &  20.33 & 6.455 & 3.464 & -0.324 & -1.463 \\
$c_{E_8}$ &  & & & & & & & & & 2.071 & 1.041 & -0.158 & -0.462 \\
$c_{E_9}$ &  & & & & & & & & & & 50.23 &  9.133 & 8.813 \\
$c_{E_{10}}$ & & & & & & & & & & & &  2.625  &   1.910 \\
$c_{E_{13}}$ & & & & & & & & & & & & & 2.499  \\ [4pt]
\hline
\end{tabular}
\end{center}
\end{table}
\newpage
%

%

\newpage

\newpage

\newpage

\newpage

\newpage

\newpage

\newpage

\newpage

\newpage

\newpage

\newpage

\newpage

\end{document}